\documentclass[journal=jacsat,manuscript=article]{achemso}
\usepackage[version=3]{mhchem} 
\usepackage[T1]{fontenc}       
\usepackage{booktabs}

\author{Yuning Zhang}
\altaffiliation{These authors contributed equally.}
\affiliation[McGill University]
{Department of Physics, McGill University, Montreal, QC, Canada}
\author{Yoichi Miyahara}
\altaffiliation{These authors contributed equally.}
\affiliation[McGill University]
{Department of Physics, McGill University, Montreal, QC, Canada}
\author{Nassim Derriche}
\affiliation[McGill University]
{Department of Physics, McGill University, Montreal, QC, Canada}
\author{Wayne Yang}
\affiliation[Unknown University]
{Department of Bionanoscience, Kavli Institute of Nanoscience Delft, Delft University of Technology}
\author{Khadija Yazda}
\affiliation[McGill University]
{Department of Physics, McGill University, Montreal, QC, Canada}
\author{Zezhou Liu}
\affiliation[McGill University]
{Department of Physics, McGill University, Montreal, QC, Canada}
\author{Peter Grutter}
\affiliation[McGill University]
{Department of Physics, McGill University, Montreal, QC, Canada}
\email{Grutter@physics.mcgill.ca}
\author{Walter Reisner}
\affiliation[McGill University]
{Department of Physics, McGill University, Montreal, QC, Canada}
\email{Reisner@physics.mcgill.ca}

\title[An \textsf{achemso} demo]
  {Nanopore fabrication via tip-controlled local breakdown using an atomic force microscope}

\abbreviations{NMR,UV}
\keywords{nanopore, AFM, dielectric breakdown, single molecule sensing, tip controlled local breakdown (TCLB)}

\begin{document}

\begin{tocentry}

Some journals require a graphical entry for the Table of Contents.
This should be laid out ``print ready'' so that the sizing of the
text is correct.

Inside the \texttt{tocentry} environment, the font used is Helvetica
8\,pt, as required by \emph{Journal of the American Chemical
Society}.

The surrounding frame is 9\,cm by 3.5\,cm, which is the maximum
permitted for  \emph{Journal of the American Chemical Society}
graphical table of content entries. The box will not resize if the
content is too big: instead it will overflow the edge of the box.

This box and the associated title will always be printed on a
separate page at the end of the document.

\end{tocentry}

\begin{abstract}

The dielectric breakdown approach for forming nanopores has greatly accelerated the pace of research in solid-state nanopore sensing, enabling inexpensive formation of nanopores via a bench top setup.  Here we demonstrate the potential of \emph{tip controlled} dielectric breakdown (TCLB) to fabricate pores 100$\times$ faster, with high scalability and nanometre positioning precision.  A conductive atomic force microscope (AFM) tip is brought into contact with a nitride membrane positioned above an electrolyte reservoir.  Application of a voltage pulse at the tip leads to the formation of a single nanoscale pore.  Pores are formed precisely at the tip position with a complete suppression of multiple pore formation.  In addition, our approach greatly accelerates the electric breakdown process, leading to an average pore fabrication time on the order of 10\,ms, at least 2 orders of magnitude shorter than achieved by classic dielectric breakdown approaches.  With this fast pore writing speed we can fabricate over 300 pores in half an hour on the same membrane.

\end{abstract}
\section{Introduction}

     Following successful demonstration of nanopore sequencing via engineered protein pores\cite{clarke2009continuous}, the next research frontier in nanopore physics is the development of solid-state nanopore devices with sequencing or diagnostic capability \cite{lindsay2016promises}. Solid-state pores are mechanically more robust, admit of cheaper, more scalable fabrication, have greater compatibility with CMOS semiconductor technology, possess enhanced micro/nanofluidic integration potential \cite{miles2013single} and could potentially increase sensing resolution \cite{lindsay2016promises}. Yet, despite the great interest in solid-state pore devices, approaches for fabricating solid-state pores, especially with diameters below 10\,nm, are limited, with the main challenge being a lack of scalable processes permitting integration of single solid-state pores with other nanoscale elements required for solid-sate sequencing schemes, such as transverse nanoelectrodes \cite{gierhart2008nanopore, ivanov2010dna}, surface plasmonic structures \cite{jonsson2013plasmonic, nicoli2014dna, pud2015self, belkin2015plasmonic, shi2018integrating} and micro/nanochannels \cite{zhang2015fabrication, zhang2018single, liu2018controlling, tahvildari2015integrating}. The main pore production approaches, such as milling via electron beams in a transmission electron microscope (TEM) \cite{storm2003fabrication} and focused-ion beam (FIB) \cite{lo2006fabrication, yang2011rapid, xia2018rapid}, use high energy beam etching of substrate material.  While these techniques can produce sub 10\,nm pores with nm positioning precision, they require expensive tools and lack scalability. 

In 2014 Kwok \emph{et al}\cite{kwok2014nanopore, briggs2014automated} showed that by directly applying a voltage across an insulating membrane in electrolyte solution, they could form single nanopores down to 2\,nm in size.  The applied voltage induces a high electric field across the thin membrane, so strong that it can induce dielectric breakdown, leading to pore formation. The dielectric breakdown method is fast, inexpensive and potentially highly scalable, yet it has a critical disadvantage:  the pore position is random.  When a high trans-membrane voltage is applied electric breakdown occurs at a ``weak'' location on the insulating membrane, a position determined randomly by the intrinsic inhomogeneity of the nitride film. As the pore can form anywhere on the membrane upon voltage application, the breakdown technique cannot form pores at precisely determined positions; creating multiple pores with well-defined spacing is likewise unfeasible.  This is a very problematic limitation, particularly given that many solid-state sensing and sequencing schemes requiring precise pore positioning (e.g. between transverse electrodes \cite{gierhart2008nanopore, ivanov2010dna}, carbon nanotubes \cite{jiang2010fabrication}, graphene nanoribbon \cite{saha2011dna}, or within a micro/nanofluidic channel \cite{zhang2015fabrication, zhang2018single, liu2018controlling}).   Multiple closely spaced pores show promise for translocation control \cite{pud2016mechanical, zhang2018single, liu2018controlling}.  Critically, the breakdown approach may also inadvertently produce more than one nanopore over the membrane area  \cite{carlsen2017solid, zrehen2017real, wang2018fabrication, ying2018formation}, leading to a drastic loss of signal-to-noise and inability to resolve single-molecule translocation events.  A recent variation of the breakdown approach uses a pipette tip to control voltage application \cite{arcadia2017situ}, increasing pore positioning precision to the micron scale (the pipette tip opening diameter is 2\,$\mu$m), but nanometer positioning precision is in fact required for many solid-state sequencing schemes, due to the small size of sensing elements required to interface with the pores. In addition, the pipette-tip approach does not prevent the potential formation of multiple pores over the still large (micron scale) region of voltage application.

\begin{figure}[!ht]
\centering
\includegraphics[width=0.7\linewidth]{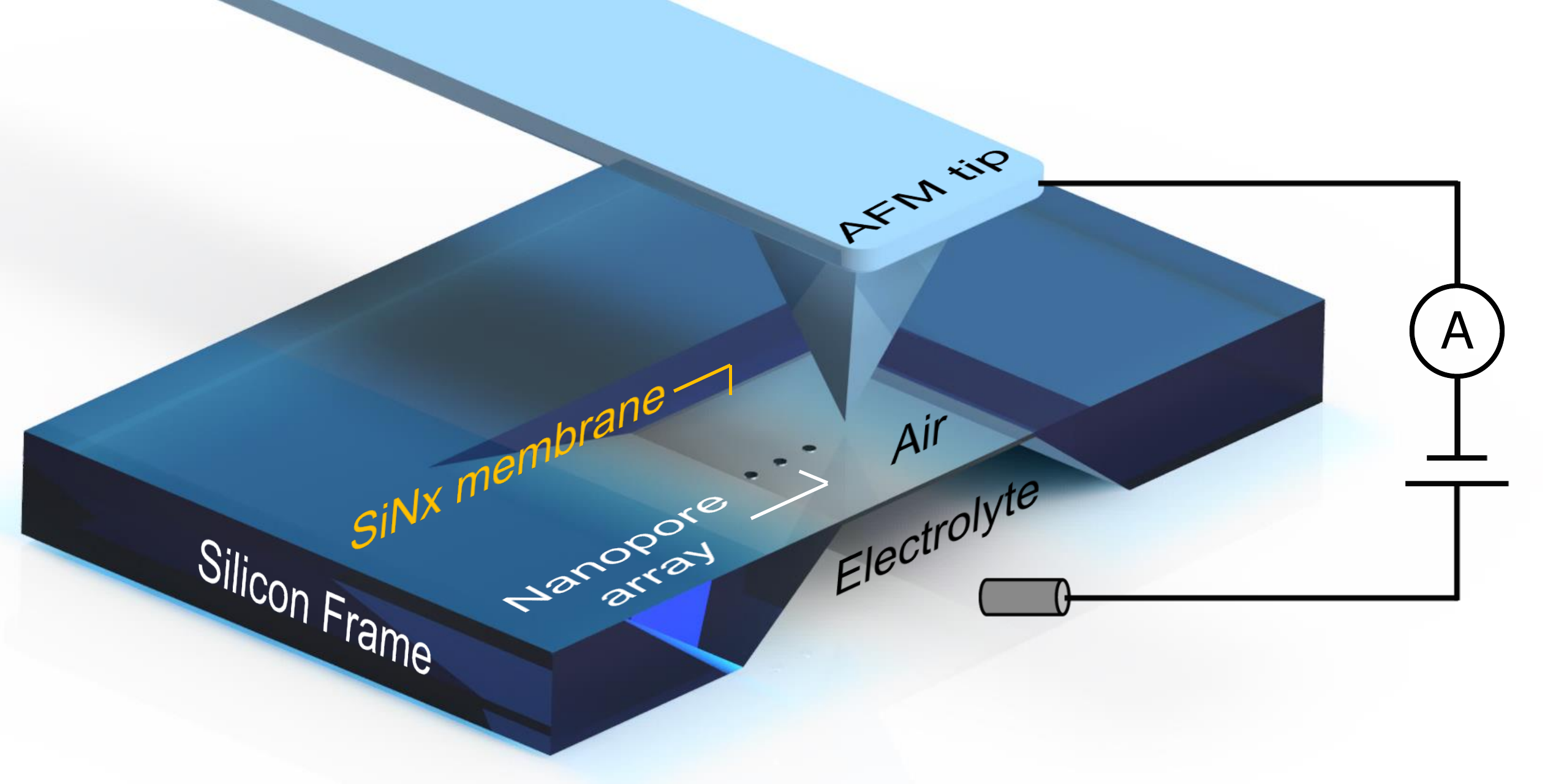}
\caption{Nanopore fabrication via tip controlled local electric breakdown (TCLB).  A 3D schematic of the experimental setup depicting an AFM cantilever with a conductive tip positioned over a silicon nitride membrane.  Application of a voltage pulse leads to formation of a nanopore at the tip position.  Nanopore arrays can be readily formed via control of the AFM tip location, with \emph{in situ} current measurement at each pore verifying successful pore fabrication at that location.  Note that our setup requires only one side of the membrane to be in contact with electrolyte, while the other side of the membrane is exposed to air.}
\label{fig:1}
\end{figure}

     We have developed a new approach for forming solid-state pores that combines the positioning advantages of particle beam milling and the simplicity/low-cost of the electric breakdown approach with the powerful imaging capabilities of Atomic Force Microscopy (AFM).  In our approach, which we call Tip-Controlled Local Breakdown (TCLB), a conductive AFM tip is brought into contact with a nitride membrane and used to apply a local voltage to the membrane (figure~\ref{fig:1}). The local voltage induces electric breakdown at a position on the membrane determined by the AFM tip, forming a nanopore at that location, which we demonstrate via I-V measurement, TEM characterization and single-molecule translocation.   Firstly, in TCLB, the nanoscale curvature of the AFM tip (r\,$\sim$\,10\,nm) localizes the electric field to a truly nanoscale region, eliminating the possibility of forming undesirable additional nanopores on the membrane as well as preventing the pore-free region of the membrane from being damaged by high electric fields.  Secondly, TCLB can form pores with a spatial precision determined by the nanoscale positioning capability of the AFM instrument (an improvement in spatial precision from micro to nanoscale).  Thirdly, TCLB drastically shortens the fabrication time of a single nanopore from on order of seconds to on order of 10\,ms (at improvement of at least 2 orders of magnitude). Fast pore fabrication implies that arrays can be written with extremely high throughput (over $\sim$100 pores in a half an hour, compared to $\sim100$ in a day \cite{arcadia2017situ}).  Fourthly, as TCLB is AFM based, it can harness the topographic, chemical and electrostatic scanning modalities of an AFM to image the membrane before and after pore formation, enabling precise alignment of pores to existing features.  The scanning capabilities of the AFM tool can be used to automate fabrication of arrays of precisely positioned pores, with the successful fabrication of each pore automatically verified by current measurement at the tip following voltage application.  The precise control of the contact force, made possible by AFM, is essential for establishing the reliable contact between the tip and the membrane. As AFM are benchtop tools that operate in ambient conditions (e.g. at atmospheric pressure and normal indoor humidity) they are inherently low-cost and can be readily scaled.  The ability to work in ambient conditions implies that the approach is compatible with materials possessing sensitive which require chemical functionalization (e.g. that might be damaged by vacuum conditions used in FIB and TEM).  Finally, while classic dielectric breakdown requires that both sides of the membrane be in contact with aqueous electrolyte reservoirs, our approach requires that only one side of the membrane be in contact with a liquid reservoir, considerably easing the scaling of our method and the speed of nanopore formation, as the AFM scanning takes place in a dry environment.  

\section{Results}

\subsection{Nanopore Fabrication}
\begin{figure}[!ht]
\centering
\includegraphics[width=0.7\linewidth]{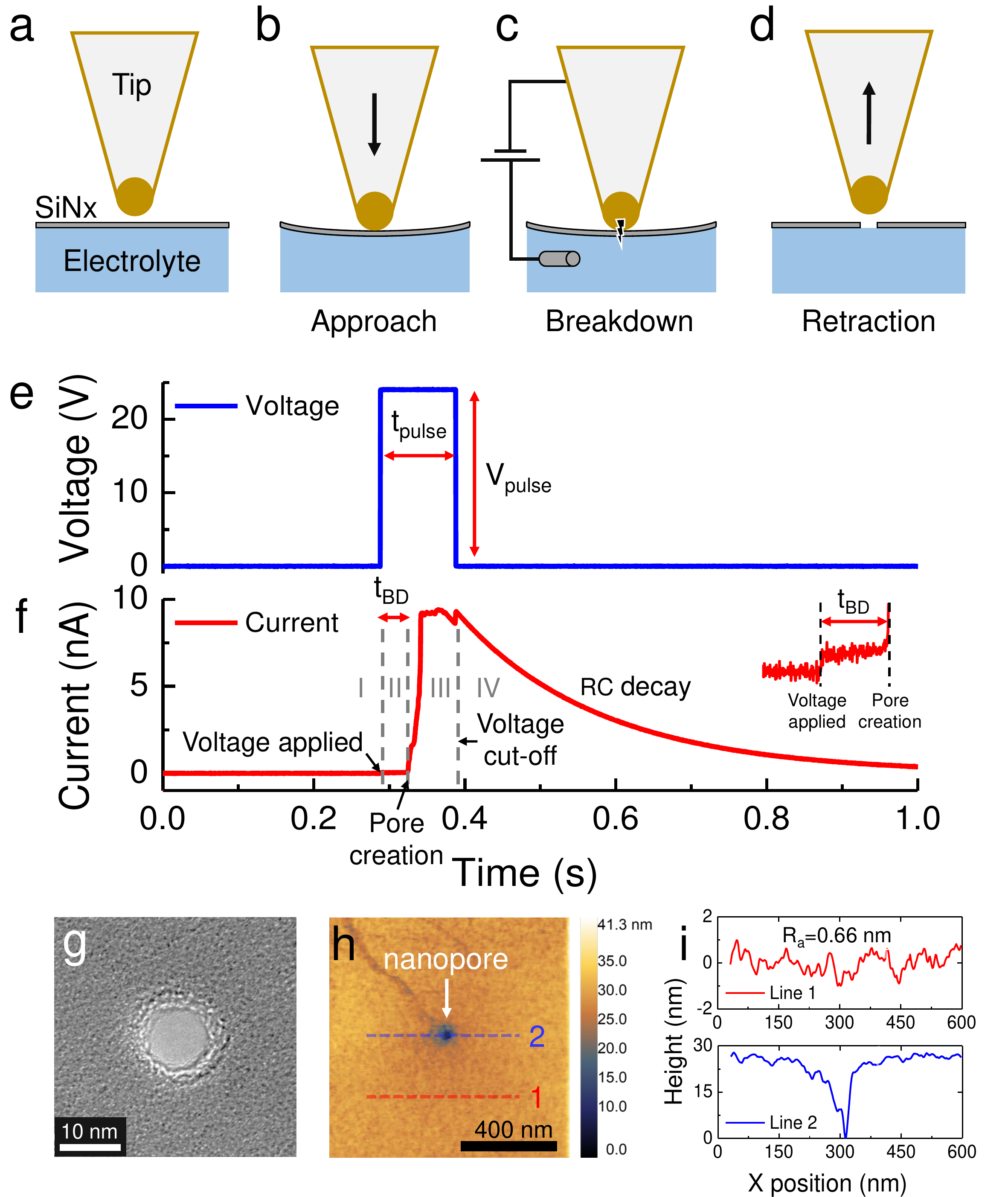}
\caption{Fabrication process of a single nanopore by conductive tip induced local electric breakdown. (a) Schematic showing the conductive AFM tip located above a thin nitride membrane. The bottom side of the membrane is in contact with electrolyte. (b) To minimize contact resistance between the tip and membrane, the tip is pressed against the membrane in contact mode. (c) A voltage pulse is applied across the membrane through the tip, initiating the breakdown process, resulting in the formation of a single nanopore. (d) Tip is retracted from the membrane once a nanopore is formed. The \textbf{voltage pulse} (e) and the \textbf{current} across the membrane (f) during a typical nanopore fabrication event.  The membrane thickness is 20\,nm, the pulse height $V_\text{pulse}=24$\,V, the pulse width $t_\text{pulse}=100$\,ms and the tip radius is $10 \pm 5$\,nm. (g)  TEM image of a 9.2\,nm pore corresponding to the current and voltage trace shown in e-f.  (h) AFM scan of a larger sized single nanopore fabricated on nitride membrane using TCLB with accompanying topographic scans of bare membrane (i-red, surface roughness $Ra=0.66$\,nm) and across the pore (i-blue). Note that small nanopores (d$\le$10 nm) may not show up on an AFM scan due to the tip radius.}
\label{fig:2}
\end{figure}

 The schematic of the experimental setup is illustrated in figure~\ref{fig:1}.  Using a bench-top AFM setup operated in ambient laboratory conditions, a conductive AFM tip is brought into contact with a thin nitride membrane sitting on top of an electrolyte reservoir.  The conductive AFM tip is positioned a distance of $\sim$\,100\,$\mu$m from the membrane (figure \ref{fig:2} a). To initiate pore fabrication, the tip approaches the membrane at a speed of $\sim$5\,$\mu$m/s until it engages with the surface (figure~\ref{fig:2} b).   A small loading force (typically in the order of 1\,nN) is applied to the tip in order to minimize contact resistance between the tip and the membrane. This force is set sufficiently small to avoid tip-induced mechanical damage to the membrane.   To initiate the breakdown process, the tip is positioned at the desired location in the scanning region and a single rectangular pulse is applied (figure~\ref{fig:2} d). The pulse has an amplitude of $V_\text{pulse}$, and a duration of $t_\text{pulse}$.  The applied voltage pulse initiates the breakdown process and creates a nanoscale pore on the membrane, located at the tip location.  After nanopore formation, the tip is retracted from the membrane (figure~\ref{fig:2} d).   A representative breakdown event is shown in figure~\ref{fig:2} e-g.  A voltage pulse of $V_\text{pulse}$=24\,V, $t_\text{pulse}$=100\,ms is applied (figure~\ref{fig:2} e). After voltage application, the current increases to $\sim$\,50\,pA and remains roughly constant (figure~\ref{fig:2} f inset). After a time delay of $t_\text{BD}$=36.2\,ms (figure~\ref{fig:2} f), the current increases sharply to a few nA, indicating successful breakdown and nanopore formation. If the pores are large, successful nanopore fabrication at the tip location can additionally be confirmed by a subsequent topographic AFM scan (figure~\ref{fig:2} h,i). When the nanopore diameter is smaller or comparable (d$\le$10\,nm) to the tip radius of curvature, the nanopore may not be observed in the AFM scan. 
 
  We have developed a custom script enabling automatic control of the pore fabrication process.  Using this script we can readily create pore arrays, iterating the single-pore formation process over a $5\times5$ grid with the pores spaced evenly by 500\,nm. Using the same tip, we have successfully fabricated over 300 nanopores on the same membrane, demonstrating the scalability of our TCLB technique (see supplementary S2 for more information).
  
\subsection{Probing the breakdown threshold}

\begin{figure}[!ht]
\centering
\includegraphics[width=0.8\linewidth]{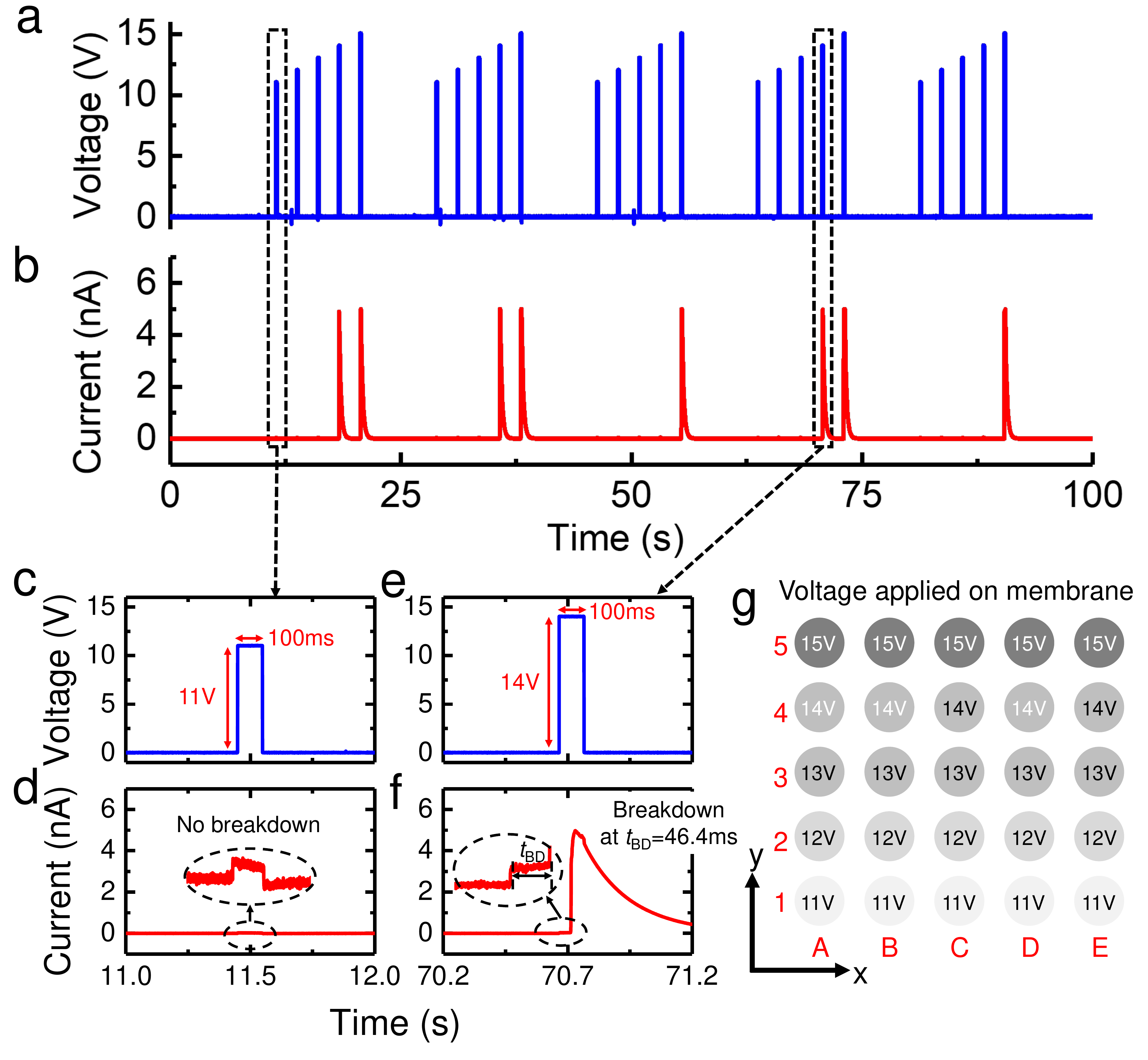}
\caption{Automatic probing of pore fabrication conditions. (a) The voltage pulse train applied to different membrane locations and (b) the resulting current.  (c-d) Magnified view of voltage pulse (c) and resulting current (d) that does not correspond to pore fabrication.  (e-f) Magnified view of voltage pulse (e) and resulting current (f) that does correspond to pore fabrication.  (g) Pore formation conditions across the 5$\times$5 array.  Location A1 corresponds to (c-d); Location D4 corresponds to (e-f).}
\label{fig:3}
\end{figure}

    Our automated pore fabrication protocol enables efficient varying of process parameters to optimize pore fabrication.  In particular, we vary the pulse amplitude across the nanopore array to probe the threshold at which membrane breakdown occurs.  A pulse train of five subsets, with each set containing five rectangular pulses of fixed duration (100\,ms) but increasing amplitude (11\,V to 15\,V, with an increment of 1\,V), are applied across the membrane (figure~\ref{fig:3} a, blue trace).  Each pulse is applied to a different location on the membrane.  The detected current is shown in Figure~\ref{fig:3} b (trace in red).  The locations are arrayed spatially in a $5\times5$ square grid, with the pulse location in the array given by figure~\ref{fig:3} g.  The fabrication process starts from location A1 and ends at location E5, rastering in the $y$ direction (figure~\ref{fig:3} g, A1$\rightarrow$A5, B1$\rightarrow$B5, C1$\rightarrow$C5, D1$\rightarrow$D5, E1$\rightarrow$E5).  The spacing between each fabrication site is 500\,nm. Spikes in the detected current, which occur for pulse amplitudes greater than $13$\,V, indicate successful electric breakdown.    At $V_\text{pulse}$=14\,V, 2 out of 5 attempts induce breakdown.   A further increase of the voltage to 15\,V leads to a 100\% breakdown probability (5 out of 5).  Magnified view of no-breakdown and successful breakdown events are shown in figure~\ref{fig:3} (c-f) corresponding to location A1 ($V_\text{pulse}$=11\,V) and D4 ($V_\text{pulse}$=14\,V).  
    
\subsection{TEM characterization}

\begin{figure}[!ht]
\centering
\includegraphics[width=0.8\linewidth]{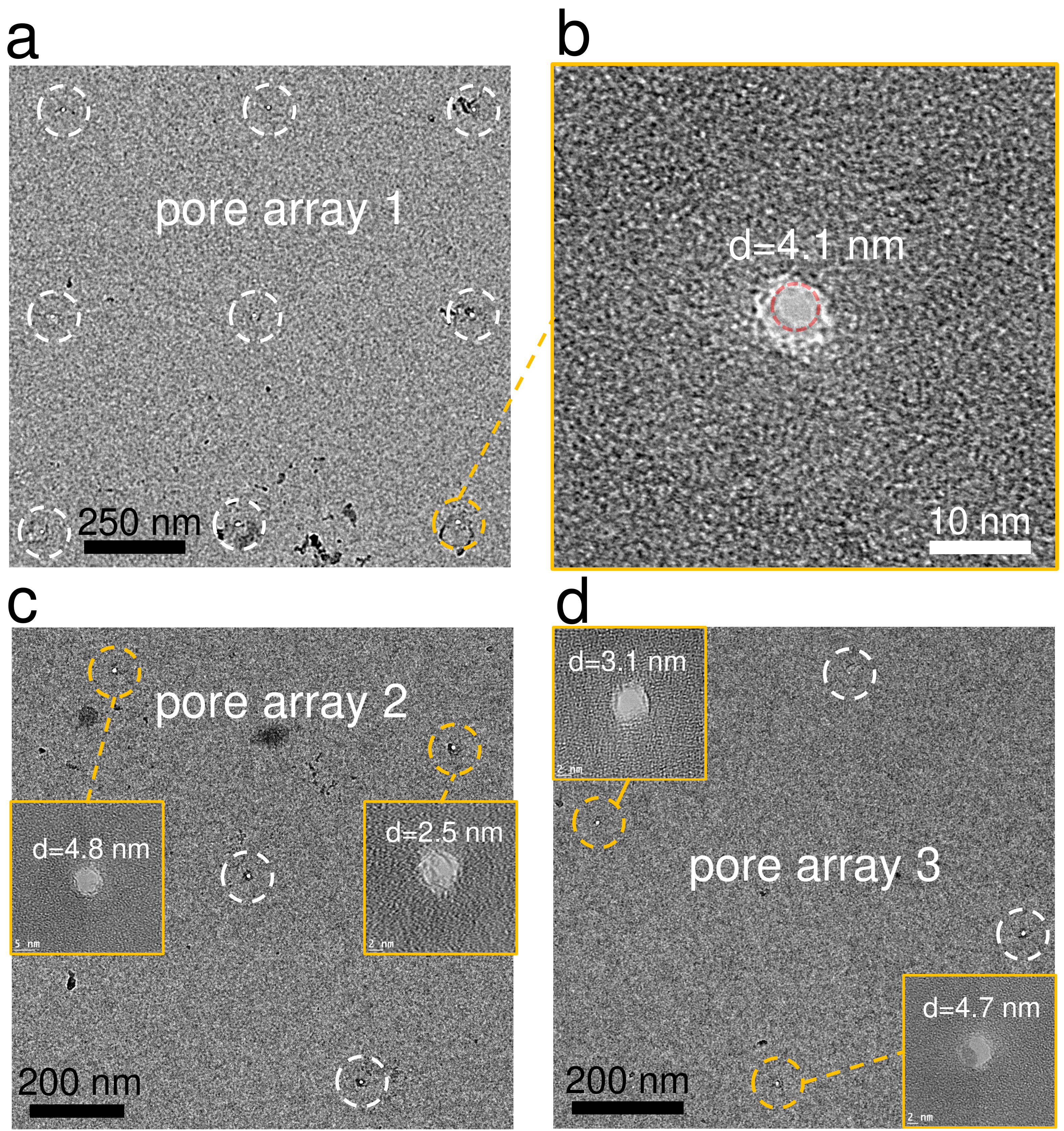}
\caption{TEM characterization of nanopore arrays. (a) TEM micrograph of a nanopore array containing 9 nanopores. Nanopores are located at the center of the dashed circles. The pore-to-pore spacing is $\sim$500\,nm.  (b) Zoomed-in TEM micrograph of a nanopore with an opening diameter of 4.1\,nm. (c)-(d) TEM micrograph of nanopore arrays fabricated on a different membrane from (a). Insets showing magnified micrographs of different nanopores with diameter close to or under 5\,nm. Fabrication condition: $V_\text{pulse}=15$\,V, $t_\text{pulse}=100$\,ms, membrane thickness $l=$12-14\,nm, tip radius $r=10\pm5$\,nm. Additional examples of nanopore arrays are shown in supplementary figure~S3.}
\label{fig:4}
\end{figure}

TEM microscopy allowed for a detailed characterization of the nanopores made by TCLB.  Figure~\ref{fig:4} shows three TEM micrographs of nanopore arrays. In agreement with our AFM settings (figure~\ref{fig:2} j and figure~\ref{fig:3} g), nanopores are spaced evenly by 500\,nm in an array format. Figure~\ref{fig:4} a and b show a 3$\times$3 nanopore array fabricated using $V_\text{pulse}$= 15\,V, $t_\text{pulse}$= 100\,ms.  Figure~\ref{fig:4} c and d show two nanopore arrays made on a new membrane with a new tip under exactly the same fabrication conditions ($V_\text{pulse}$= 15\,V, $t_\text{pulse}$= 100\,ms). Despite using different tips and membranes (12-14\,nm thick) from different chips, nanopores fabricated with the same parameters as our TCLB method have similar diameters (below or close to 5\,nm).

\subsection{Pore Formation Mechanism}
\subsubsection{Weibull versus Log-normal}
Nanopore fabrication time (time-to-breakdown, $t_\text{BD}$) can provide insight into the pore formation mechanism. Nanopores fabricated via classic dielectric breakdown have a time-to-breakdown following a Weibull probability distribution \cite{briggs2015kinetics, arcadia2017situ, yanagi2018two}. The Weibull distribution is used extensively to model the time-to-failure of semiconductor devices \cite{dissado1984weibull, degraeve1995consistent}. The Weibull distribution arises from the ``weakest-link'' nature of typical dielectric breakdown process, where breakdown happens at the weakest position over a large membrane area.  The nanopore fabrication time is dominated by the time to make a pore at this weakest position.

\begin{figure}[!ht]
\centering
\includegraphics[width=0.9\linewidth]{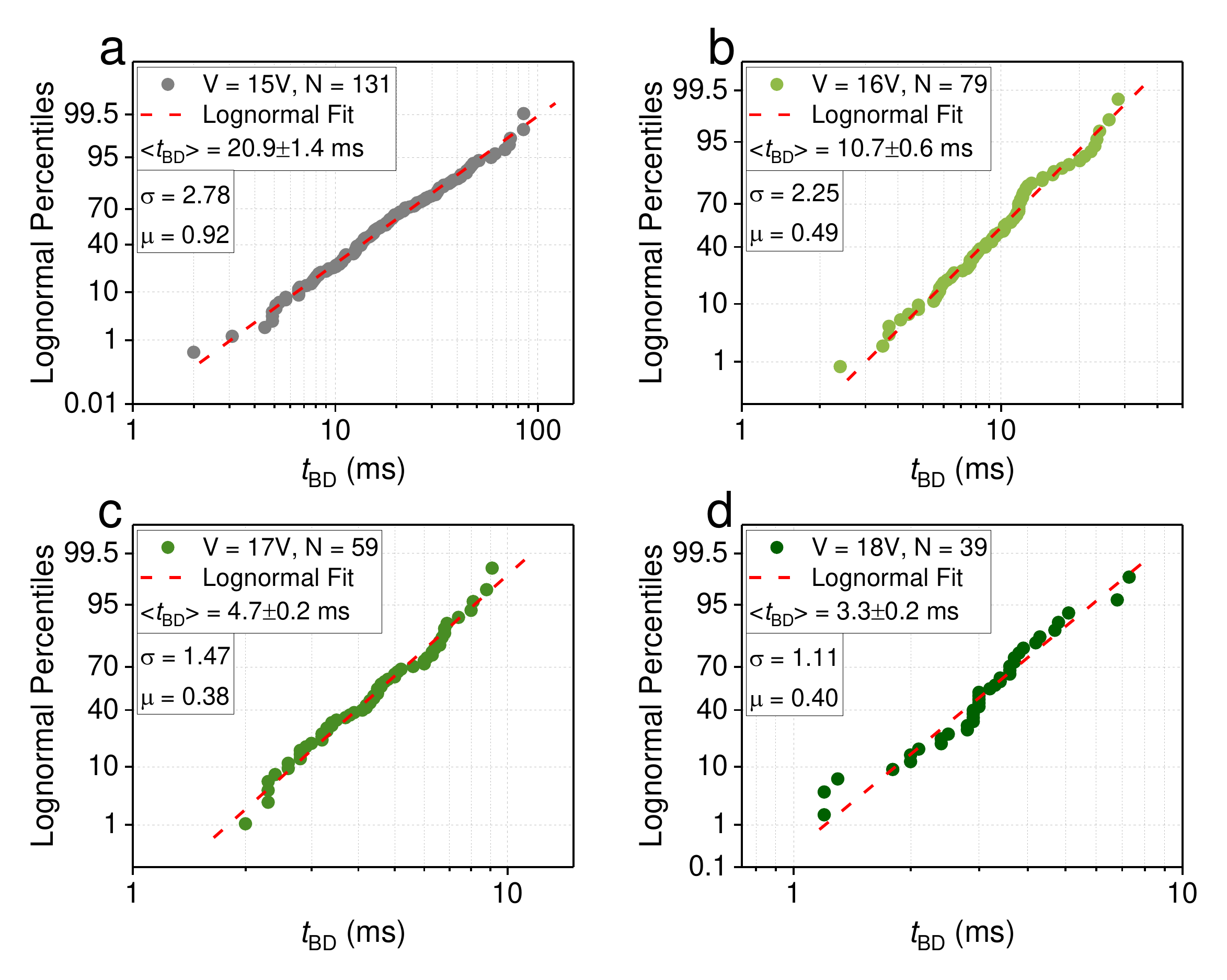}
\caption{Log-normal probability plot of time-to-breakdown ($t_\text{BD}$) for a total of 308 nanopores under different pulse voltages. (a) Cumulative distribution of $t_\text{BD}$ presented with a log-normal rescaling under following conditions: $V_\text{pulse}=$15\,V, $t_\text{pulse}=$100\,ms, membrane thickness $l = $12--14\,nm. The average nanopore fabrication time is <$t_\text{BD}$>$=20.9\pm 1.4$\,ms. (b)--(d) Cumulative distributions of $t_\text{BD}$ with $V_\text{pulse}=$16, 17, 18\,V  respectively. The dashed lines give the best fit to a log-normal distribution.  All experiments are performed with the same tip on one membrane. Tip radius of curvature: $\sim$10\,nm. Membrane thickness: 12--14\,nm. Window size: 50$\times$50\,$\mu$m$^{2}$. (See supplementary section S4 for more details regarding log-normal distribution, Weibull distribution and probability plots.)}
\label{fig:5}
\end{figure}

In contrast, we find that our time-to-breakdown distribution, obtained from forming over 300 nanopores using our automatic process, yields better agreement with a \emph{log-normal} probability distribution.  Figure \ref{fig:5} shows the cumulative distribution of time-to-breakdown plotted with a log-normal scaling.  In this form,  data distributed according to a log-normal distribution follows a straight line.  Our time-to-breakdown results, linearized by this rescaling, are thus clearly consistent with a log-normal distribution.   In figure \ref{fig:S4}, we plot the same results rescaled appropriately for a Weibull, and it is apparent that the Weibull is not as good a description.  See supplementary materials section 4 for more detail on log-normal, Weibull distribution and appropriate rescalings (probability plot forms).  

The better agreement with a log-normal suggests that the physical mechanism of pore-formation is different using TCLB than classic breakdown.  Under tip control, the membrane location where dielectric breakdown occurs is controlled by the tip position, and is thus highly defined rather than random.  In this case the statistics of membrane breakdown is no longer a weakest link problem (i.e. determined by the time to breakdown of some randomly located ``weak-point''), but instead is determined by the degradation of a ``typical'' location on the membrane reflecting average film properties. Theoretical and experimental work demonstrate that the overall time-scale of a degradation process that arises from the multiplicative action of many small degradation steps (regardless of physical mechanism) can be modelled via a log-normal distribution \cite{peck1974reliability, berman1981time, lloyd2005simple, mcpherson2010reliability}. Possible degradation mechanisms for our pore-formation process include electromigration, diffusion and corrosion \cite{strong2009reliability}.

\subsubsection{Voltage dependence of time-to-breakdown}

\begin{figure}[!ht]
\centering
\includegraphics[width=0.9\linewidth]{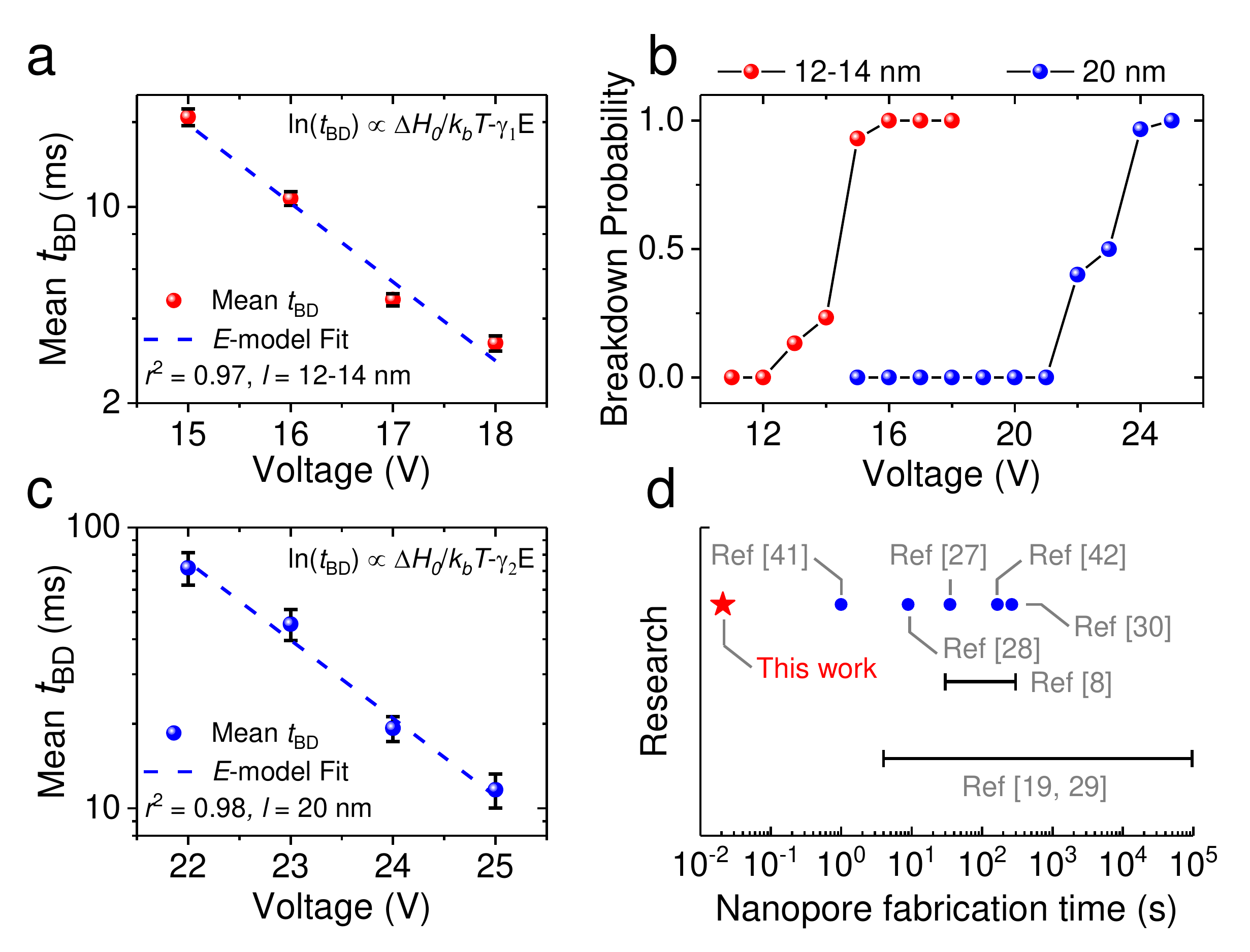}
\caption{(a) Semilog plot of the mean breakdown time ($\langle t_\text{BD} \rangle$) versus voltage for 12-14\,nm thick nitride membrane with an exponential fit.  (b)  Breakdown probability versus voltage for 12-14\,nm and 20\,nm thick nitride.  (c)  Semilog plot of the mean breakdown time versus voltage for a 20\,nm thick nitride membrane.  (d)  Comparison of average nanopore fabrication time of this work versus range of studies exploring classical breakdown \cite{kwok2014nanopore, briggs2015kinetics, yanagi2014fabricating, pud2015self, arcadia2017situ, ying2018formation, yanagi2018two, yamazaki2018photothermally}. Note that if the average fabrication time is not given or can not be estimated from the reference, a range is then plotted for comparison (see more details in supplementary section S5).}
\label{fig:6}
\end{figure}

In figure \ref{fig:6} a we show the mean time-to-breakdown ($\langle t_\text{BD} \rangle$) versus voltage on a semi-log scale.  The mean time-to-breakdown decreases exponentially with voltage.  This behaviour is predicted by the \emph{E}-model of time dependent dielectric breakdown (TDDB) \cite{mcpherson1998underlying}, which predicts that the mean time-to-breakdown should depend exponentially on the local electric field (proportional to applied voltage at the tip).  The \emph{E}-model arises fundamentally from a thermochemical \cite{mcpherson1998underlying, mcpherson2003thermochemical} rather than a direct tunnelling mechanism (Fowler-Nordheim tunnelling) \cite{mcpherson1998comparison}.  In thermochemical breakdown, high voltage across the dielectric material induces strong dipolar coupling of local electric field with intrinsic defects in the dielectric. Weak bonding states can be thermally broken due to this strong dipole-field coupling, which in turn serves to lower the activation energy required for thermal bond-breakage and accelerates the degradation process, resulting in a final dielectric breakdown \cite{mcpherson1998underlying, mcpherson2003thermochemical}.

We have also investigated whether we can use tip-controlled breakdown to produce pores in thicker (20\,nm) nitride membranes.   We are able to form pores with a high probability but with a corresponding increase in the required voltage, as demonstrated by figure 6b.  The mean time-to-breakdown as a function of voltage in the thicker membranes also follow the \emph{E}-model (figure \ref{fig:6} c).

In figure 6d we compare the average time-to-breakdown for our tip controlled approach versus classical dielectric breakdown.  We find that our approach gives pore formation times two orders of magnitude lower than classical breakdown, by comparison with a wide-range of experimental studies \cite{kwok2014nanopore, briggs2015kinetics, yanagi2014fabricating, pud2015self, arcadia2017situ, ying2018formation, yanagi2018two, yamazaki2018photothermally, bandara2019push} exploring classical breakdown for different film thickness (10-30\,nm, 75\,nm), pH (2-13.5) and voltage (1-24\,V). 

\subsection{Single Molecule DNA Detection}
\begin{figure}[!ht]
\centering
\includegraphics[width=0.8\linewidth]{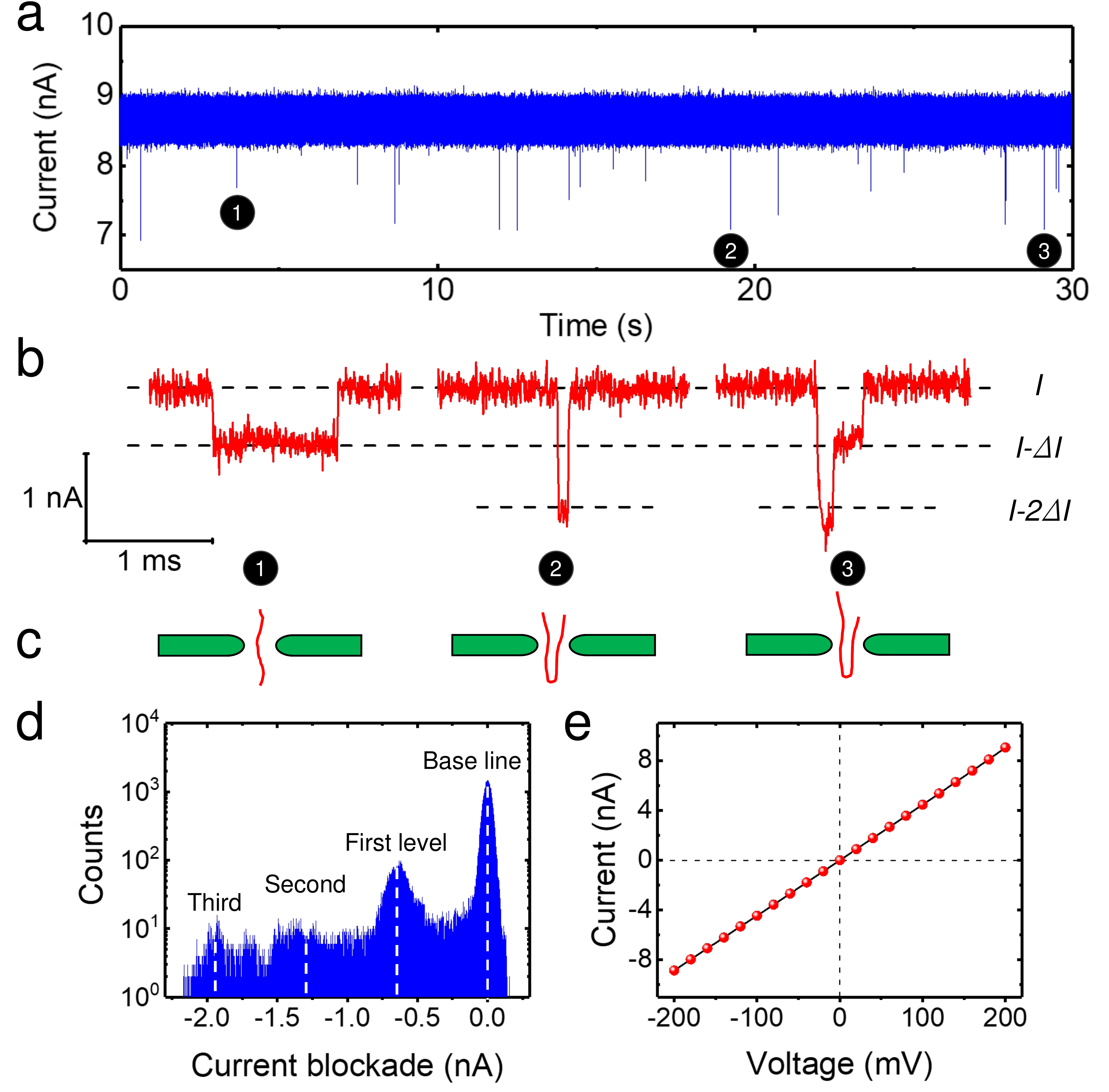}
\caption{DNA translocation through a nanopore fabricated using TCLB ($V_\text{pulse}$=20\,V, $t_\text{pulse}$=150\,ms, membrane thickness 10\,nm, tip radius $r$=10$\pm$5\,nm). (a) Typical ionic current traces of DNA translocating through a 9.9\,nm pore in a 10\,nm thick nitride membrane. Experiment was performed with 0.5\,$\mu$g/mL 100\,bp ladder DNA (100-2000\,bp) in 1\,M KCl buffered with 10\,mM Tris, 1\,mM EDTA, at pH=8.0.  Observed events are labelled as 1-3, corresponding to different DNA configurations/folding states while translocating through the pore. (b) Zoomed-in current trace of event 1, 2 and 3, corresponding to the cartoon translocation types shown in (c).  (d) Current blockade histogram including over 500 events. (e) I-V characterization of the nanopore. The nanopore displays an Ohmic I-V curve with a resistance of 23.0\,M$\Omega$, leading to an effective pore diameter of 9.9\,nm. Power spectral density (PSD) of the nanopore is shown in supplementary figure \ref{fig:S5}. }
\label{fig:7}
\end{figure}

Lastly, we show nanopores produced using our tip-controlled approach can be used for single molecule detection.  Figure \ref{fig:7} shows results for 100\,bp ladder DNA (100-2000\,bp) translocating through a 9.9\,nm pore ($V_\text{pulse}$=20\,V, $t_\text{pulse}$=150\,ms, membrane thickness 10\,nm, tip radius $r$=10$\pm$5\,nm). To perform single molecule detection, the chip is transferred to a fluidic cell with DNA containing 1\,M KCl buffer added to the $cis$ chamber and DNA-free buffer added to the $trans$ chamber.  A potential drop of 200\,mV is applied across the nanopore, so that DNA molecule are pulled from $cis$ to $trans$ through the pore. 
Figure \ref{fig:7} a-b shows typical signatures of ionic blockades induced by translocating DNA, composed of a mixture of single and multi-level events. A histogram of current blockades, including 587 translocation events measured by the same nanopore, is shown in Figure \ref{fig:7} d. Prior to performing this DNA translocation experiment, an I-V trace was obtained to characterize pore size (figure \ref{fig:7} e), which yielded a nanopore resistance of 23.0\,M$\Omega$. This strong linearity between current and applied voltage demonstrates that our TCLB fabricated nanopore has an outstanding Ohmic performance. Using a membrane thickness $l$=10\,nm and an electrolyte conductivity $\sigma$=10\,S/m, according to the pore conductance model\cite{kowalczyk2011modeling} the estimated effective pore diameter is 9.9\,nm. 

\section{Discussion and Conclusion}

     In summary, we show that tip-controlled local breakdown can be used to produce pores with nm positioning precision (determined by AFM tip), high scalability (100's of pores over a single membrane) and fast formation (100$\times$ faster than classic breakdown) using a bench-top tool.  These capabilities will greatly accelerate the field of solid-state nanopore research.  In particular, the nm positioning is crucial for wide-range sensing and sequencing applications where there is a need to interface nanopores with additional nanoscale elements.   Sequencing approaches based on tunneling require positioning a pore between two electrodes \cite{gierhart2008nanopore, ivanov2010dna}.  Plasmonic devices with interfaced pores require positioning pores at the optimal distance ($10-20$\,nm) from nano antennas in order to maximize plasmonic coupling \cite{jonsson2013plasmonic, nicoli2014dna, pud2015self, belkin2015plasmonic, shi2018integrating}.  In devices utilizing nanofluidic confinement (e.g. nanochannels, nanocavities) pores need to be aligned with etched sub 100\,nm features \cite{zhang2015fabrication, zhang2018single, liu2018controlling, larkin2017length}.  In addition to producing pores, our AFM based approach can exploit multiple scanning modalities (topographic, chemical, electrostatic) to map the device prior to pore production and so align pores precisely to existing features.   
     
    TCLB can be integrated into an automated wafer-scale AFM system, ensuring nm alignment of each pore with simultaneous mass pore production.  Thus, not only can TCLB drive novel nanopore sensing applications, TCLB can simultaneously drive the industrial scaling of these applications.  As an example, consider combining TCLB with photo-thermally assisted thinning\cite{yamazaki2018photothermally, gilboa2018optically, ying2018formation}.   In a photo-thermally assisted thinning process, a laser beam is focused on a silicon nitride membrane, leading to formation of a locally thinned out region, with thinning achieved down to a few nm \cite{yamazaki2018photothermally}.  If there is only one thinned well formed, classic dielectric breakdown will tend to form a pore at this `thinned out'  weakest position.  Classic dielectric breakdown, however, is limited to forming only \emph{one} pore in \emph{one} well across an entire membrane.  In contrast, TCLB can position pores in each member of a large-scale array of photo-thermally thinned wells, with the wells packed as close as the photo-thermal thinning technique allows.  Specifically, AFM topographic scans will determine the center-point of each well and TCLB will then form pores at these positions.
   
   TCLB may also have applications beyond nanopore fabrication, providing an AFM-based approach to locally characterize the dielectric strength of thin membranes and 2D materials.  This application, which could be useful for the MEMS and the semiconductor industry, could enable mapping of dielectric strength across large membranes and semiconductor devices, leading to enhanced material performance (e.g. for high-$\kappa$ gate dielectrics \cite{okada2007dielectric}).


\section{Methods}
\indent \textbf{Materials.} 
The nitride membranes we use are commercially available from Norcada (part \# NBPT005YZ-HR and NT002Y). The membrane is supported by a circular silicon frame (2.7\,$\mu$m diameter, 200\,$\mu$m thickness) with a window size of 10$\times$10, 20$\times$20 or 50$\times$50\,$\mu$m$^{2}$.  The membrane thickness is 10\,nm, 12-14\,nm or 20\,nm. 
The AFM probes used are obtained from Adama Innovations (part \# AD-2.8-AS) and have a tip radii of curvature of 10$\pm$5\,nm. 
Nanopore fabrication experiments are performed in 1\,M sodium percholorate dissolved in propylene carbonate (PC), with a conductivity of 2.82\,S/m \cite{daprano1996conductance}. DNA translocation experiments are performed in a 3D printed fluidic cell with 100\,bp ladder DNA (Sigma-Aldrich, 100-2000\,bp) diluted to a final concentration of 0.5\,$\mu$g/mL in 1\,M KCl buffered with 10\,mM Tris and 1\,mM EDTA at pH=8.0.

\noindent \textbf{Instrumentation.} The atomic force microscope used in our experiments is a MultiMode Nanoscope III from Digital Instruments (now Bruker). Nanoscript is used for automated fabrication of nanopores. The TEM images are acquired using the JEM-2100F TEM from JEOL.

\noindent \textbf{Current Data Acquisition and Analysis.} The current signal during nanopore fabrication is recorded using a custom current amplifier with 5\,kHz detection bandwidth.  Analysis of dielectric breakdown events in the current signal was performed using a custom Python code.  The ionic trans-pore current during DNA translocations was recorded using an Axopatch 200B with a 250\,kHz sampling rate, low-pass filtered at 100\,kHz. DNA translocation data analysis was carried out using Transalyzer\cite{plesa2015data}.

\begin{acknowledgement}
This work is financially supported by the Natural Sciences and Engineering Research Council of Canada (NSERC) Discovery Grants Program (Grant No. RGPIN 386212 and RGPIN 05033), Idea to Innovation (I2I) Grant (I2IPJ 520635-18) and joint CIHR funded Canadian Health Research Projects grant (CIHRR CPG-140199).  The authors acknowledge useful discussions with Prof. Robert Sladek and Hooman Hosseinkhannazer. The authors acknowledge Norcada for material supplies (nitride membranes). The authors acknowledge Facility for Electron Microscopy Research (FEMR) at McGill and Centre de Caract\'erisation Microscopique des Mat\'eriaux (CM)$^{2}$ at \,Ecole Polytechnique de Montr\'eal for access to electron microscopes.
\end{acknowledgement}

\newpage
\begin{suppinfo}

\renewcommand{\thefigure}{S\arabic{figure}}
\setcounter{figure}{0}
\renewcommand{\thetable}{S\arabic{table}}
\setcounter{table}{0}
\renewcommand{\theequation}{S\arabic{equation}}
\setcounter{equation}{0}

\subsection{S1-Experimental Setup}
Here we discuss the detailed experimental setup for TCLB nanopore fabrication. The fluidic cell assembly and accompanying schematic are shown in figure \ref{fig:S1} a and b. Prior to pore fabrication, the circular nitride TEM window is mounted in the fluidic cell with the cell body filled by electrolyte. The cell is then placed inside the AFM headstage (figure \ref{fig:S1} c).  Alignment of the conductive AFM tip to the nitride membrane is monitored via two optical microscopes with an external light source.

\begin{figure}[!ht]
\centering
\includegraphics[width=0.8\linewidth]{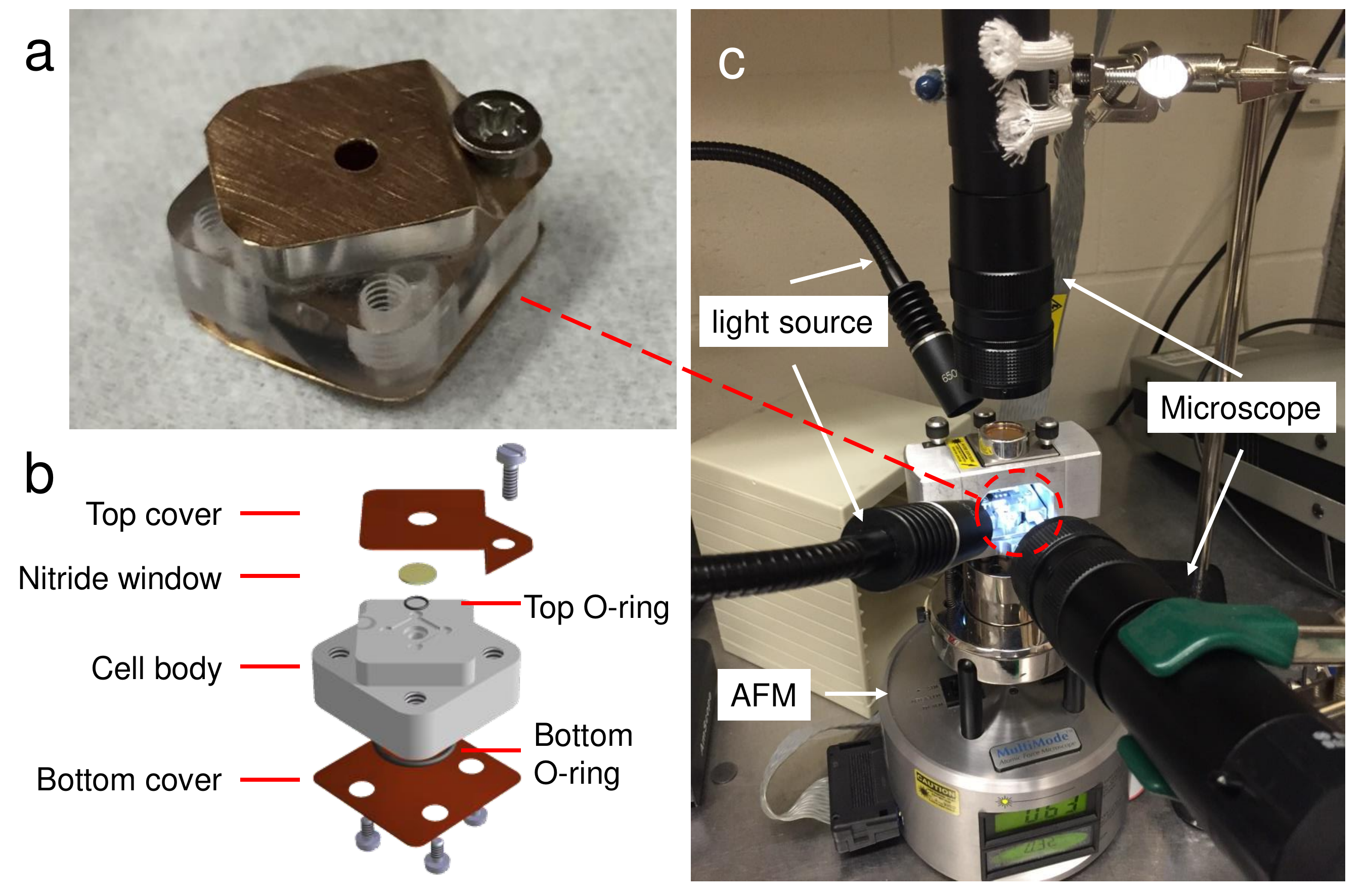}
\caption{Experimental setup for nanopore fabrication. (a) Assembled fluidic cell with nitride window mounted. (b) Fluidic cell assembly with nitride windows  sandwiched between the top cover and O-ring. (c) AFM setup.  The whole setup is mounted on a vibration isolation table.}
\label{fig:S1}
\end{figure}

\newpage
\subsection{S2-Reliability and scalability of TCLB}
To demonstrate the reliability and scalability of the TCLB technique, we fabricated over 300 nanopores using the same AFM tip on one membrane. All data presented in figure \ref{fig:5} are collected from a total of 308 nanopores, fabricated using a single tip on the same membrane window (12-14\,nm thick, window size 50$\times$50\,$\mu$m$^{2}$) with a total time of around 30\,min.   The location of the arrays (17 in total) in relation to the window position are mapped out in figure \ref{fig:S2} a. Each array contains a maximum possible of 25 nanopores (5$\times$5 array). Figure \ref{fig:S2} b shows another example of 11 arrays (in total 68 nanopores) located on a 20\,nm thick membrane, window size 20$\times$20\,$\mu$m$^{2}$.

\begin{figure}[!ht]
\centering
\includegraphics[width=1\linewidth]{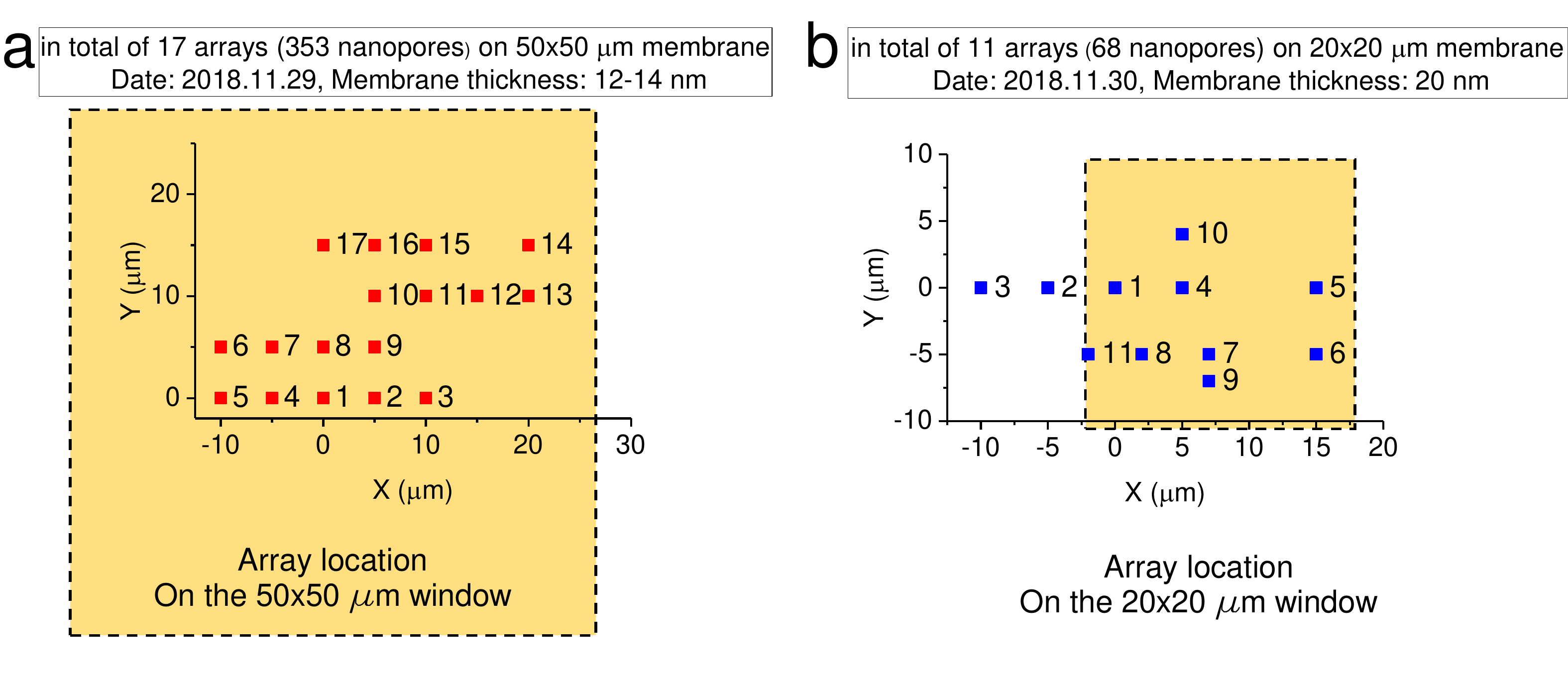}
\caption{Mapping the location of nanopore arrays across the membrane. (a) Schematic showing the position of 17 nanopore arrays on a 50$\times$50\,$\mu$m membrane window (12-14\,nm thick). Each array contains a maximum possible 25 nanopores (5$\times$5 array).  A total of 353 nanopores were successfully fabricated under various fabrication conditions. (a) The position of 11 nanopore arrays on a 20$\times$20\,$\mu$m membrane window (20\,nm thick).  A total of 68 nanopores were successfully fabricated. }
\label{fig:S2}
\end{figure}

\newpage
\subsection{S3-TEM characterization}
Two additional TEM micrographs of nanopore arrays are presented in figure \ref{fig:S3}.

\begin{figure}[!ht]
\centering
\includegraphics[width=0.7\linewidth]{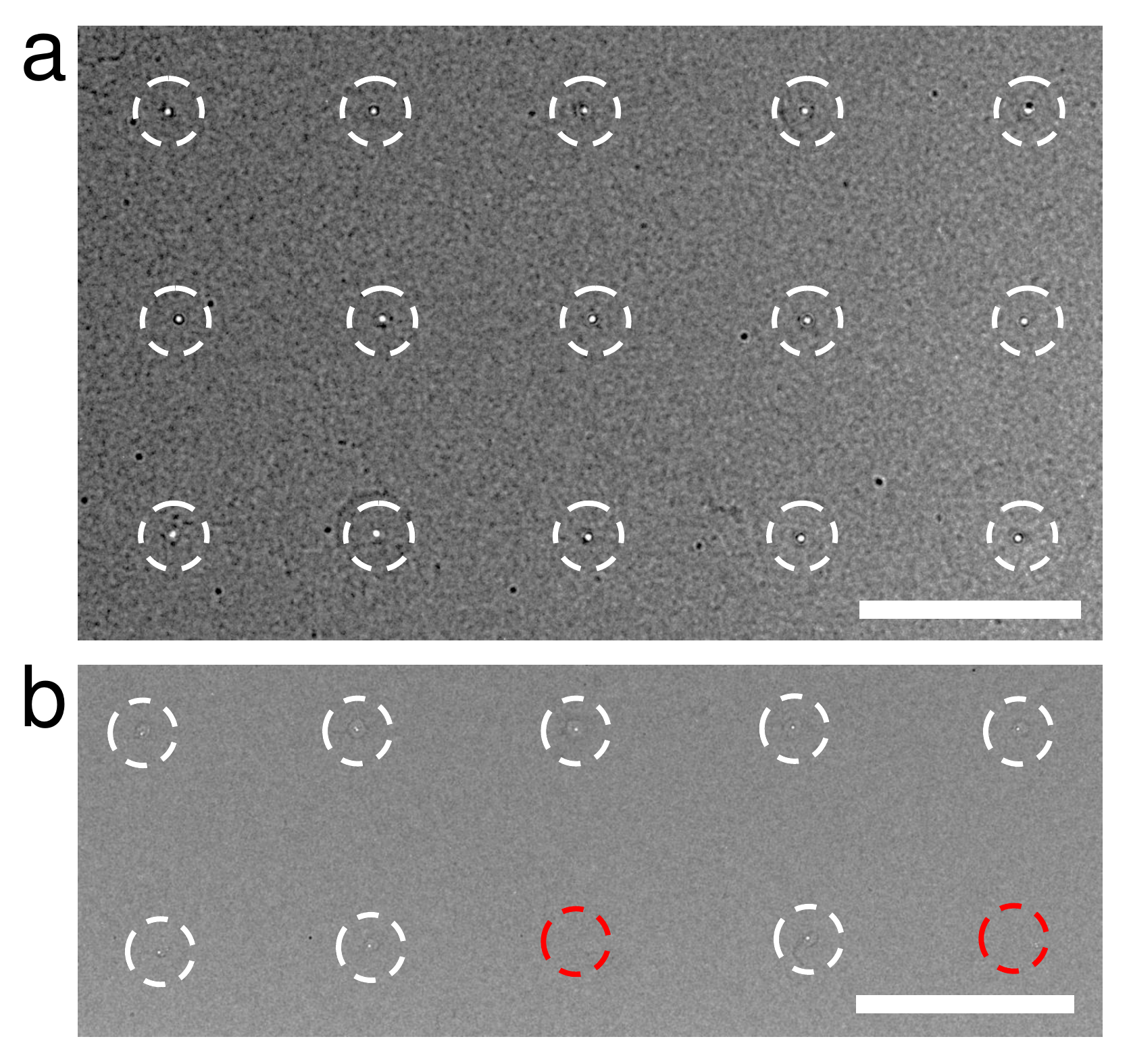}
\caption{TEM micrograph of nanopore arrays.  (a) TEM micrograph showing a 3$\times$5 nanopore array fabricated on a 20\,nm nitride membrane. All nanopore are fabricated using the same conditions: $V_\text{pulse}$=25\,V, $t_\text{pulse}$=100\,ms. (b)  TEM micrograph showing a 2$\times$5 nanopore array fabricated on a 20\,nm nitride membrane. The pores in the top row are fabricated using $V_\text{pulse}$=24\,V and $t_\text{pulse}$=100\,ms. The yield is 100\% (5 out of 5). The pores in the bottom row are fabricated using the same pulse width ($t_\text{pulse}$=100\,ms) but lower voltage ($V_\text{pulse}$=23\,V). The yield is only 60\% (red circles indicate failed breakdown attempts). The scale bar is 500\,nm. Note that TEM images are taken slightly under focused in order to better visualize the nanopore.}
\label{fig:S3}
\end{figure}

\newpage
\subsection{S4-Log-normal distribution and Weibull distribution}
The probability density function (\textit{pdf}) and cumulative distribution function (\textit{cdf}) of \emph{log-normal} distribution and 2-parameter \emph{Weibull} distribution are given by:\\
\text{Log-normal \textit{pdf}:} 
\begin{equation}    
f(t; \mu, \sigma)=\frac{1}{t \sigma \sqrt{2 \pi}} e^{-\frac{(\ln t- \mu)^2}{2 {\sigma}^2}}
\label{Lognormal:pdf}
\end{equation}
\text{Log-normal \textit{cdf}:} 
\begin{equation}
F(t; \mu, \sigma)= \frac{1}{2}+\frac{1}{2} \text{erf}[\frac{\ln t-\mu}{\sqrt{2}\sigma}]
\label{Lognormal:cdf}
\end{equation}
\text{Weibull \textit{pdf}:} 
\begin{equation}
f(t; \lambda, \beta)=\frac{\beta}{\lambda} (\frac{t}{\lambda})^{\beta-1} e^{-{(\frac{t}{\lambda})}^{\beta}}
\label{Weibull:pdf}
\end{equation}
\text{Weibull \textit{cdf}:}
\begin{equation}
F(t; \lambda, \beta)=1- e^{-(\frac{t}{\lambda})^{\beta}}
\label{Weibull:cdf}
\end{equation}
where $\mu$, $\sigma$ are log-normal distribution's shape and scale parameters; $\beta$ and $\lambda$ are Weibull distribution's shape and scale parameters. The symbol erf designates the error function, $\text{erf}(x)=\frac{2}{\sqrt{\pi}} \int_{0}^{x} e^{-t^{2}}dt$. \emph{F}(\emph{t}) is the cumulative failure rate at time \emph{t} (t$\ge$0).
\subsubsection*{Probability plots of log-normal and Weibull}
The probability plots of distributions are constructed by rescaling the axes to linearize the cumulative distribution function (\emph{cdf}) of the distribution. For example after rescaling both X axis and Y axis, a log-normal distribution will show up as a straight line in the log-normal probability plot, likewise a Weibull distribution will show up as a straight line in the Weibull probability plot.
The X scale type and Y scale type for log-normal probability plot and Weibull probability plot are given by:

\begin{table}[]
\begin{tabular}{lll}
Distribution & X scale type & Y scale type          \\ \hline
Log-normal   & Ln           & Probability           \\ \hline
Weibull      & Log10        & Double Log Reciprocal
\end{tabular}

\label{probabilityplots}
\end{table}

\noindent where \emph{Probability} scaling is given by the inverse of a cumulative Gaussian distribution: $X^{-1}=\Phi^{-1}(X/100)$. The quantity $\Phi$ is the cumulative Gaussian distribution function, $\Phi=\frac{1}{2} [1+\text{erf}(\frac{x-\mu}{\sigma \sqrt{2}})]$. \emph{Double log reciprocal} scaling is given by $X^{-1}=\ln(-\ln(1-X))$.  

An example of the log-normal probability plot of time-to-breakdown ($t_\text{BD}$) is shown in figure \ref{fig:5}. An example of the Weibull probability plot for the same data set is shown in figure \ref{fig:S4}. One can compare probability plots for log-normal and Weilbull and conclude that the time-to-breakdown ($t_\text{BD}$) fits better to a log-normal distribution.

\begin{figure}[!ht]
\centering
\includegraphics[width=0.9\linewidth]{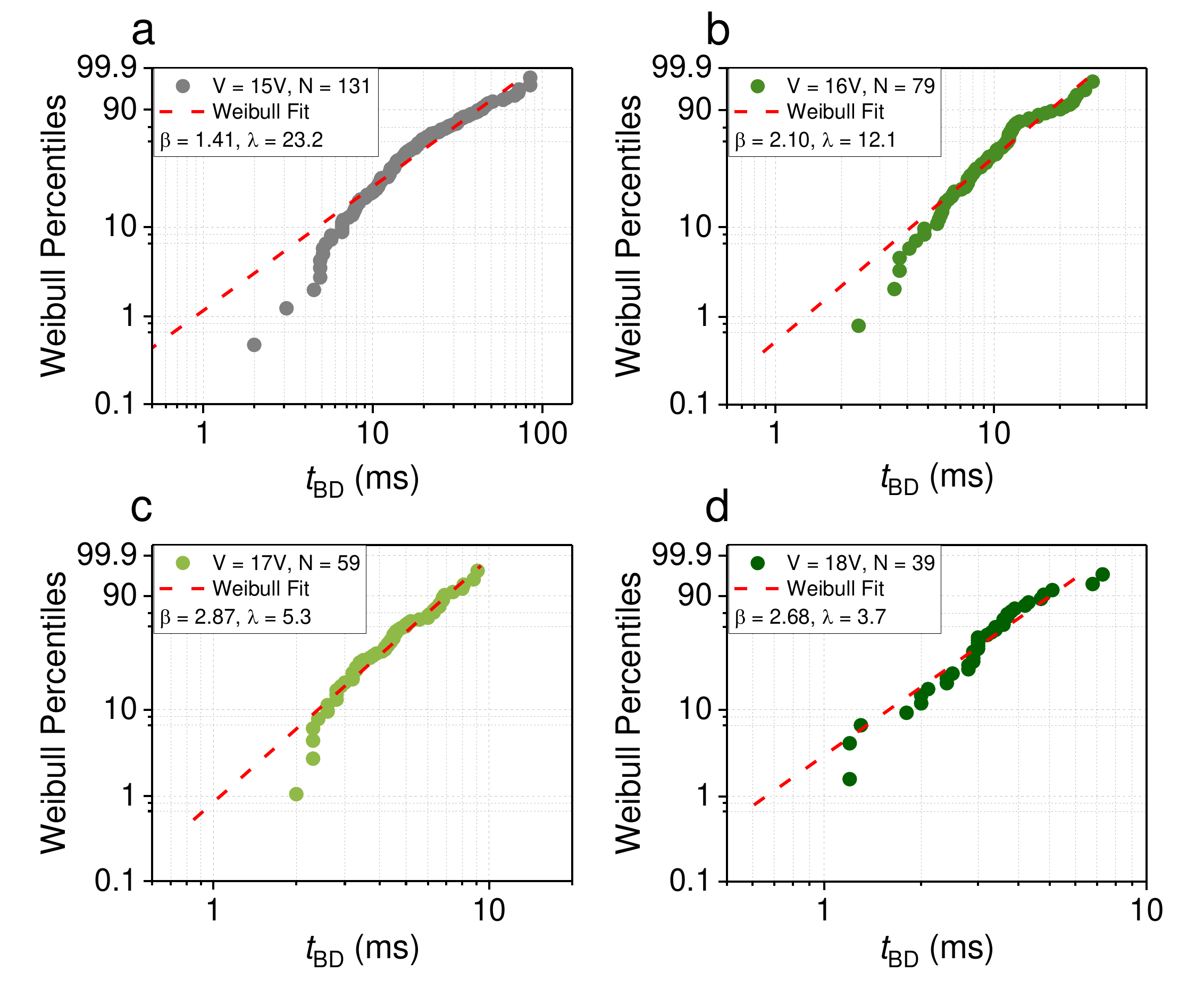}
\caption{Weibull probability plot of time-to-breakdown ($t_\text{BD}$) for the same data set presented in figure~\ref{fig:5}. (a) Cumulative distribution of $t_\text{BD}$ presented with a Weibull rescaling under following conditions: $V_\text{pulse}=$15\,V, $t_\text{pulse}=$100\,ms, membrane thickness $l = $12--14\,nm. The average nanopore fabrication time is $\langle t_\text{BD} \rangle$=20.9$\pm$1.4\,ms. (b)--(d) Cumulative distributions of $t_\text{BD}$ with $V_\text{pulse}=$16, 17, 18\,V  respectively. The dashed lines give the best fit to a Weibull distribution. All experiments are performed with the same tip on one membrane. Tip radius of curvature: $\sim$10\,nm.}
\label{fig:S4}
\end{figure}

\newpage
\subsection{S5-Nanopore fabrication time comparison}
The following table compares dielectric breakdown based nanopore fabrication approaches in greater detail, including average pore fabrication time, membrane thickness, breakdown voltage, min/max fabrication time and number of nanopores analyzed.

\begin{table}[]
\small\addtolength{\tabcolsep}{-4.5pt}
\begin{tabular}{ccccccc}
\hline
Methods                                                                         & \begin{tabular}[c]{@{}c@{}}Average pore\\ fabrication time\end{tabular} & \begin{tabular}[c]{@{}c@{}}Membrane\\ thickness\end{tabular}                             & \begin{tabular}[c]{@{}c@{}}Breakdown\\ voltage\end{tabular} & \multicolumn{2}{c}{\begin{tabular}[c]{@{}c@{}}Min/Max\\ fabrication time\end{tabular}} & \begin{tabular}[c]{@{}c@{}}Number of \\ nanopores analyzed\end{tabular} \\ \hline
{\color[HTML]{000000} This work}                                                & {\color[HTML]{000000} 20\,ms}                                            & {\color[HTML]{000000} \begin{tabular}[c]{@{}c@{}}10\,nm, 12-14\,nm, \\ 20\,nm\end{tabular}} & {\color[HTML]{000000} 13-25\,V}                              & {\color[HTML]{000000} 1\,ms}               & {\color[HTML]{000000} 85\,ms}               & {\color[HTML]{000000} $\sim$400}                                        \\ \hline
CBD \cite{kwok2014nanopore, briggs2015kinetics}                                                                    & NA                                                                      & 10\,nm, 30\,nm                                                                             & 5-17\,V                                                      & 4\,s                                       & 10$^{5}$\,s                    & $\sim$50                                                                \\ \hline
Micro pipette \cite{arcadia2017situ}                                                           & 8.9\,s                                                                   & 10\,nm                                                                                    & up to 24\,V                                                  & 1\,s                                       & 17\,s                                       & 169                                                                     \\ \hline
Two-step BD \cite{yanagi2018two}                                                           & 265.5\,s                                                                 & 20\,nm                                                                                    & 10, 20\,V                                                    & 150\,s                                     & 350\,s                                      & 50                                                                      \\ \hline
\begin{tabular}[c]{@{}c@{}}Multilevel pulse\\ injection \cite{yanagi2014fabricating} \end{tabular}    & $\sim$1\,s                                                               & 10\,nm                                                                                    & 2.5, 7\,V                                                     & 0.1\,s                                     & 20\,s                                       & 40                                                                      \\ \hline
\begin{tabular}[c]{@{}c@{}}Optically \\ controlled BD \cite{pud2015self} \end{tabular}      & NA                                                                      & 20\,nm                                                                                    & 6\,V                                                         & 30\,s                                      & 300\,s                                      & NA                                                                      \\ \hline
\begin{tabular}[c]{@{}c@{}}Laser-assisted \\ controlled BD \cite{ying2018formation} \end{tabular} & $\sim$35\,s                                                              & 30\,nm                                                                                    & 18\,V                                                        & 10\,s                                      & 80\,s                                       & 33 \\ \hline
\begin{tabular}[c]{@{}c@{}}Photothermally\\ assisted BD \cite{yamazaki2018photothermally} \end{tabular}    & 165\,s                                                                   & 75\,nm                                                                                    & 1\,V                                                         & NA                                        & NA                                         & 29                                                                     \\ \hline
\end{tabular}
\caption{Comparing nanopore fabrication of TCLB with classical dielectric breakdown.}
\label{table:S1}
\end{table}

\newpage
\subsection{S6-Nanopore PSD}
Figure below shows the current power spectral density (PSD) plot of the nanopore presented in figure \ref{fig:7}.

\begin{figure}[!ht]
\centering
\includegraphics[width=0.6\linewidth]{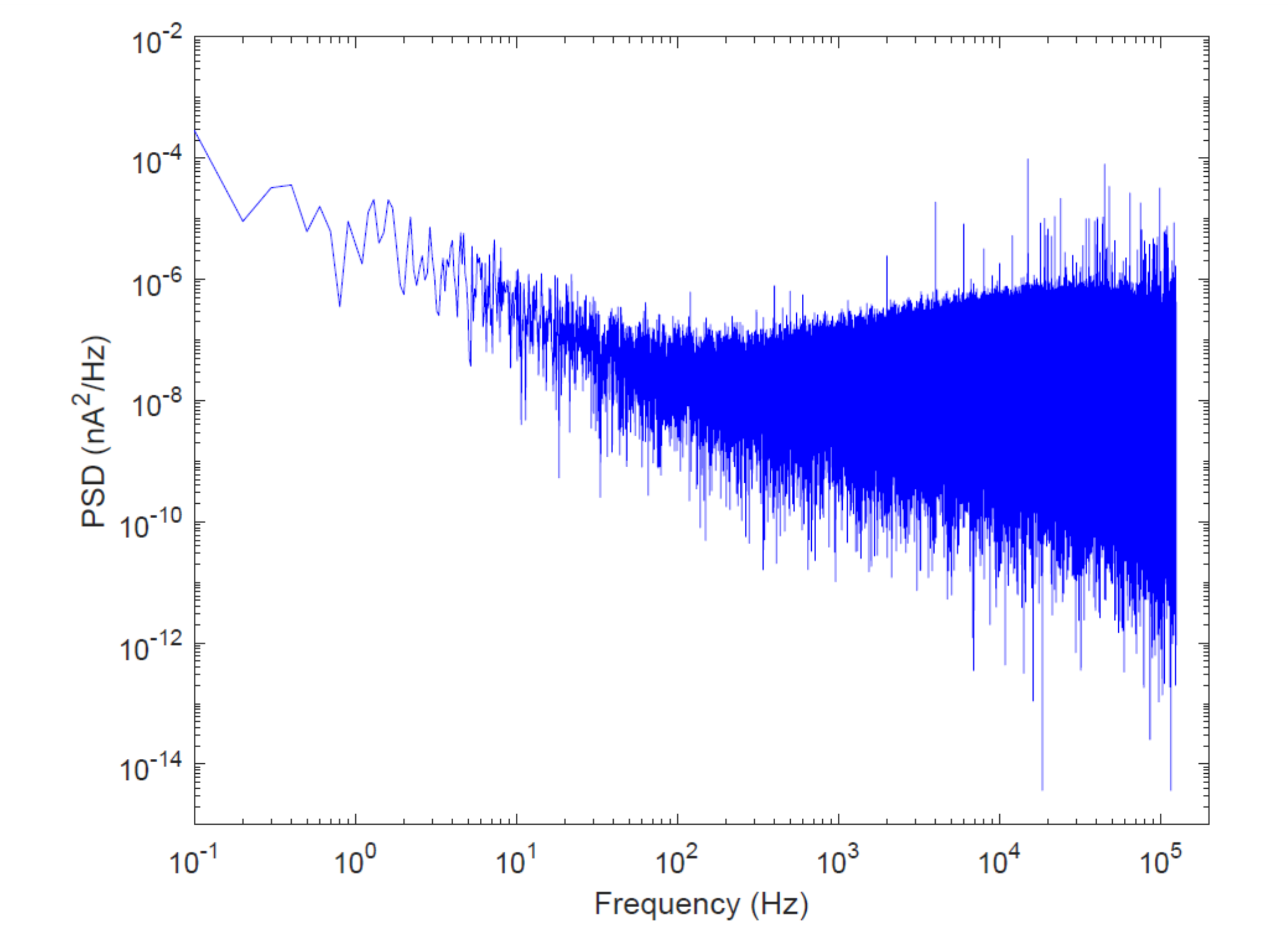}
\caption{Current power spectral density (PSD) of the nanopore presented in figure \ref{fig:7}. PSD was obtained by measuring a 10\,s of ionic trace, low-pass filtered at 100\,kHz and sampled at 250\,kHz with 200\,mV applied voltage.}
\label{fig:S5}
\end{figure}

\end{suppinfo}

\newpage
\bibliography{AFMnanopore.bib}

\providecommand{\latin}[1]{#1}
\makeatletter
\providecommand{\doi}
  {\begingroup\let\do\@makeother\dospecials
  \catcode`\{=1 \catcode`\}=2 \doi@aux}
\providecommand{\doi@aux}[1]{\endgroup\texttt{#1}}
\makeatother
\providecommand*\mcitethebibliography{\thebibliography}
\csname @ifundefined\endcsname{endmcitethebibliography}
  {\let\endmcitethebibliography\endthebibliography}{}
\begin{mcitethebibliography}{50}
\providecommand*\natexlab[1]{#1}
\providecommand*\mciteSetBstSublistMode[1]{}
\providecommand*\mciteSetBstMaxWidthForm[2]{}
\providecommand*\mciteBstWouldAddEndPuncttrue
  {\def\EndOfBibitem{\unskip.}}
\providecommand*\mciteBstWouldAddEndPunctfalse
  {\let\EndOfBibitem\relax}
\providecommand*\mciteSetBstMidEndSepPunct[3]{}
\providecommand*\mciteSetBstSublistLabelBeginEnd[3]{}
\providecommand*\EndOfBibitem{}
\mciteSetBstSublistMode{f}
\mciteSetBstMaxWidthForm{subitem}{(\alph{mcitesubitemcount})}
\mciteSetBstSublistLabelBeginEnd
  {\mcitemaxwidthsubitemform\space}
  {\relax}
  {\relax}

\bibitem[Clarke \latin{et~al.}(2009)Clarke, Wu, Jayasinghe, Patel, Reid, and
  Bayley]{clarke2009continuous}
Clarke,~J.; Wu,~H.-C.; Jayasinghe,~L.; Patel,~A.; Reid,~S.; Bayley,~H.
  \emph{Nature nanotechnology} \textbf{2009}, \emph{4}, 265\relax
\mciteBstWouldAddEndPuncttrue
\mciteSetBstMidEndSepPunct{\mcitedefaultmidpunct}
{\mcitedefaultendpunct}{\mcitedefaultseppunct}\relax
\EndOfBibitem
\bibitem[Lindsay(2016)]{lindsay2016promises}
Lindsay,~S. \emph{Nature nanotechnology} \textbf{2016}, \emph{11}, 109\relax
\mciteBstWouldAddEndPuncttrue
\mciteSetBstMidEndSepPunct{\mcitedefaultmidpunct}
{\mcitedefaultendpunct}{\mcitedefaultseppunct}\relax
\EndOfBibitem
\bibitem[Miles \latin{et~al.}(2013)Miles, Ivanov, Wilson, Do{\u{g}}an, Japrung,
  and Edel]{miles2013single}
Miles,~B.~N.; Ivanov,~A.~P.; Wilson,~K.~A.; Do{\u{g}}an,~F.; Japrung,~D.;
  Edel,~J.~B. \emph{Chemical Society Reviews} \textbf{2013}, \emph{42},
  15--28\relax
\mciteBstWouldAddEndPuncttrue
\mciteSetBstMidEndSepPunct{\mcitedefaultmidpunct}
{\mcitedefaultendpunct}{\mcitedefaultseppunct}\relax
\EndOfBibitem
\bibitem[Gierhart \latin{et~al.}(2008)Gierhart, Howitt, Chen, Zhu, Kotecki,
  Smith, and Collins]{gierhart2008nanopore}
Gierhart,~B.~C.; Howitt,~D.~G.; Chen,~S.~J.; Zhu,~Z.; Kotecki,~D.~E.;
  Smith,~R.~L.; Collins,~S.~D. \emph{Sensors and Actuators B: Chemical}
  \textbf{2008}, \emph{132}, 593--600\relax
\mciteBstWouldAddEndPuncttrue
\mciteSetBstMidEndSepPunct{\mcitedefaultmidpunct}
{\mcitedefaultendpunct}{\mcitedefaultseppunct}\relax
\EndOfBibitem
\bibitem[Ivanov \latin{et~al.}(2010)Ivanov, Instuli, McGilvery, Baldwin,
  McComb, Albrecht, and Edel]{ivanov2010dna}
Ivanov,~A.~P.; Instuli,~E.; McGilvery,~C.~M.; Baldwin,~G.; McComb,~D.~W.;
  Albrecht,~T.; Edel,~J.~B. \emph{Nano letters} \textbf{2010}, \emph{11},
  279--285\relax
\mciteBstWouldAddEndPuncttrue
\mciteSetBstMidEndSepPunct{\mcitedefaultmidpunct}
{\mcitedefaultendpunct}{\mcitedefaultseppunct}\relax
\EndOfBibitem
\bibitem[Jonsson and Dekker(2013)Jonsson, and Dekker]{jonsson2013plasmonic}
Jonsson,~M.~P.; Dekker,~C. \emph{Nano letters} \textbf{2013}, \emph{13},
  1029--1033\relax
\mciteBstWouldAddEndPuncttrue
\mciteSetBstMidEndSepPunct{\mcitedefaultmidpunct}
{\mcitedefaultendpunct}{\mcitedefaultseppunct}\relax
\EndOfBibitem
\bibitem[Nicoli \latin{et~al.}(2014)Nicoli, Verschueren, Klein, Dekker, and
  Jonsson]{nicoli2014dna}
Nicoli,~F.; Verschueren,~D.; Klein,~M.; Dekker,~C.; Jonsson,~M.~P. \emph{Nano
  letters} \textbf{2014}, \emph{14}, 6917--6925\relax
\mciteBstWouldAddEndPuncttrue
\mciteSetBstMidEndSepPunct{\mcitedefaultmidpunct}
{\mcitedefaultendpunct}{\mcitedefaultseppunct}\relax
\EndOfBibitem
\bibitem[Pud \latin{et~al.}(2015)Pud, Verschueren, Vukovic, Plesa, Jonsson, and
  Dekker]{pud2015self}
Pud,~S.; Verschueren,~D.; Vukovic,~N.; Plesa,~C.; Jonsson,~M.~P.; Dekker,~C.
  \emph{Nano letters} \textbf{2015}, \emph{15}, 7112--7117\relax
\mciteBstWouldAddEndPuncttrue
\mciteSetBstMidEndSepPunct{\mcitedefaultmidpunct}
{\mcitedefaultendpunct}{\mcitedefaultseppunct}\relax
\EndOfBibitem
\bibitem[Belkin \latin{et~al.}(2015)Belkin, Chao, Jonsson, Dekker, and
  Aksimentiev]{belkin2015plasmonic}
Belkin,~M.; Chao,~S.-H.; Jonsson,~M.~P.; Dekker,~C.; Aksimentiev,~A. \emph{ACS
  nano} \textbf{2015}, \emph{9}, 10598--10611\relax
\mciteBstWouldAddEndPuncttrue
\mciteSetBstMidEndSepPunct{\mcitedefaultmidpunct}
{\mcitedefaultendpunct}{\mcitedefaultseppunct}\relax
\EndOfBibitem
\bibitem[Shi \latin{et~al.}(2018)Shi, Verschueren, Pud, and
  Dekker]{shi2018integrating}
Shi,~X.; Verschueren,~D.; Pud,~S.; Dekker,~C. \emph{Small} \textbf{2018},
  \emph{14}, 1703307\relax
\mciteBstWouldAddEndPuncttrue
\mciteSetBstMidEndSepPunct{\mcitedefaultmidpunct}
{\mcitedefaultendpunct}{\mcitedefaultseppunct}\relax
\EndOfBibitem
\bibitem[Zhang and Reisner(2015)Zhang, and Reisner]{zhang2015fabrication}
Zhang,~Y.; Reisner,~W. \emph{Nanotechnology} \textbf{2015}, \emph{26},
  455301\relax
\mciteBstWouldAddEndPuncttrue
\mciteSetBstMidEndSepPunct{\mcitedefaultmidpunct}
{\mcitedefaultendpunct}{\mcitedefaultseppunct}\relax
\EndOfBibitem
\bibitem[Zhang \latin{et~al.}(2018)Zhang, Liu, Zhao, Yu, Reisner, and
  Dunbar]{zhang2018single}
Zhang,~Y.; Liu,~X.; Zhao,~Y.; Yu,~J.-K.; Reisner,~W.; Dunbar,~W.~B.
  \emph{Small} \textbf{2018}, \emph{14}, 1801890\relax
\mciteBstWouldAddEndPuncttrue
\mciteSetBstMidEndSepPunct{\mcitedefaultmidpunct}
{\mcitedefaultendpunct}{\mcitedefaultseppunct}\relax
\EndOfBibitem
\bibitem[Liu \latin{et~al.}(2018)Liu, Zhang, Nagel, Reisner, and
  Dunbar]{liu2018controlling}
Liu,~X.; Zhang,~Y.; Nagel,~R.; Reisner,~W.; Dunbar,~W.~B. \emph{arXiv preprint
  arXiv:1811.11105} \textbf{2018}, \relax
\mciteBstWouldAddEndPunctfalse
\mciteSetBstMidEndSepPunct{\mcitedefaultmidpunct}
{}{\mcitedefaultseppunct}\relax
\EndOfBibitem
\bibitem[Tahvildari \latin{et~al.}(2015)Tahvildari, Beamish, Tabard-Cossa, and
  Godin]{tahvildari2015integrating}
Tahvildari,~R.; Beamish,~E.; Tabard-Cossa,~V.; Godin,~M. \emph{Lab on a Chip}
  \textbf{2015}, \emph{15}, 1407--1411\relax
\mciteBstWouldAddEndPuncttrue
\mciteSetBstMidEndSepPunct{\mcitedefaultmidpunct}
{\mcitedefaultendpunct}{\mcitedefaultseppunct}\relax
\EndOfBibitem
\bibitem[Storm \latin{et~al.}(2003)Storm, Chen, Ling, Zandbergen, and
  Dekker]{storm2003fabrication}
Storm,~A.; Chen,~J.; Ling,~X.; Zandbergen,~H.; Dekker,~C. \emph{Nature
  materials} \textbf{2003}, \emph{2}, 537\relax
\mciteBstWouldAddEndPuncttrue
\mciteSetBstMidEndSepPunct{\mcitedefaultmidpunct}
{\mcitedefaultendpunct}{\mcitedefaultseppunct}\relax
\EndOfBibitem
\bibitem[Lo \latin{et~al.}(2006)Lo, Aref, and Bezryadin]{lo2006fabrication}
Lo,~C.~J.; Aref,~T.; Bezryadin,~A. \emph{Nanotechnology} \textbf{2006},
  \emph{17}, 3264\relax
\mciteBstWouldAddEndPuncttrue
\mciteSetBstMidEndSepPunct{\mcitedefaultmidpunct}
{\mcitedefaultendpunct}{\mcitedefaultseppunct}\relax
\EndOfBibitem
\bibitem[Yang \latin{et~al.}(2011)Yang, Ferranti, Stern, Sanford, Huang, Ren,
  Qin, and Hall]{yang2011rapid}
Yang,~J.; Ferranti,~D.~C.; Stern,~L.~A.; Sanford,~C.~A.; Huang,~J.; Ren,~Z.;
  Qin,~L.-C.; Hall,~A.~R. \emph{Nanotechnology} \textbf{2011}, \emph{22},
  285310\relax
\mciteBstWouldAddEndPuncttrue
\mciteSetBstMidEndSepPunct{\mcitedefaultmidpunct}
{\mcitedefaultendpunct}{\mcitedefaultseppunct}\relax
\EndOfBibitem
\bibitem[Xia \latin{et~al.}(2018)Xia, Huynh, McVey, Kobler, Stern, Yuan, and
  Ling]{xia2018rapid}
Xia,~D.; Huynh,~C.; McVey,~S.; Kobler,~A.; Stern,~L.; Yuan,~Z.; Ling,~X.~S.
  \emph{Nanoscale} \textbf{2018}, \emph{10}, 5198--5204\relax
\mciteBstWouldAddEndPuncttrue
\mciteSetBstMidEndSepPunct{\mcitedefaultmidpunct}
{\mcitedefaultendpunct}{\mcitedefaultseppunct}\relax
\EndOfBibitem
\bibitem[Kwok \latin{et~al.}(2014)Kwok, Briggs, and
  Tabard-Cossa]{kwok2014nanopore}
Kwok,~H.; Briggs,~K.; Tabard-Cossa,~V. \emph{PloS one} \textbf{2014}, \emph{9},
  e92880\relax
\mciteBstWouldAddEndPuncttrue
\mciteSetBstMidEndSepPunct{\mcitedefaultmidpunct}
{\mcitedefaultendpunct}{\mcitedefaultseppunct}\relax
\EndOfBibitem
\bibitem[Briggs \latin{et~al.}(2014)Briggs, Kwok, and
  Tabard-Cossa]{briggs2014automated}
Briggs,~K.; Kwok,~H.; Tabard-Cossa,~V. \emph{Small} \textbf{2014}, \emph{10},
  2077--2086\relax
\mciteBstWouldAddEndPuncttrue
\mciteSetBstMidEndSepPunct{\mcitedefaultmidpunct}
{\mcitedefaultendpunct}{\mcitedefaultseppunct}\relax
\EndOfBibitem
\bibitem[Jiang \latin{et~al.}(2010)Jiang, Mihovilovic, Chan, and
  Stein]{jiang2010fabrication}
Jiang,~Z.; Mihovilovic,~M.; Chan,~J.; Stein,~D. \emph{Journal of Physics:
  Condensed Matter} \textbf{2010}, \emph{22}, 454114\relax
\mciteBstWouldAddEndPuncttrue
\mciteSetBstMidEndSepPunct{\mcitedefaultmidpunct}
{\mcitedefaultendpunct}{\mcitedefaultseppunct}\relax
\EndOfBibitem
\bibitem[Saha \latin{et~al.}(2011)Saha, Drndic, and Nikolic]{saha2011dna}
Saha,~K.~K.; Drndic,~M.; Nikolic,~B.~K. \emph{Nano letters} \textbf{2011},
  \emph{12}, 50--55\relax
\mciteBstWouldAddEndPuncttrue
\mciteSetBstMidEndSepPunct{\mcitedefaultmidpunct}
{\mcitedefaultendpunct}{\mcitedefaultseppunct}\relax
\EndOfBibitem
\bibitem[Pud \latin{et~al.}(2016)Pud, Chao, Belkin, Verschueren, Huijben, van
  Engelenburg, Dekker, and Aksimentiev]{pud2016mechanical}
Pud,~S.; Chao,~S.-H.; Belkin,~M.; Verschueren,~D.; Huijben,~T.; van
  Engelenburg,~C.; Dekker,~C.; Aksimentiev,~A. \emph{Nano letters}
  \textbf{2016}, \emph{16}, 8021--8028\relax
\mciteBstWouldAddEndPuncttrue
\mciteSetBstMidEndSepPunct{\mcitedefaultmidpunct}
{\mcitedefaultendpunct}{\mcitedefaultseppunct}\relax
\EndOfBibitem
\bibitem[Carlsen \latin{et~al.}(2017)Carlsen, Briggs, Hall, and
  Tabard-Cossa]{carlsen2017solid}
Carlsen,~A.~T.; Briggs,~K.; Hall,~A.~R.; Tabard-Cossa,~V. \emph{Nanotechnology}
  \textbf{2017}, \emph{28}, 085304\relax
\mciteBstWouldAddEndPuncttrue
\mciteSetBstMidEndSepPunct{\mcitedefaultmidpunct}
{\mcitedefaultendpunct}{\mcitedefaultseppunct}\relax
\EndOfBibitem
\bibitem[Zrehen \latin{et~al.}(2017)Zrehen, Gilboa, and Meller]{zrehen2017real}
Zrehen,~A.; Gilboa,~T.; Meller,~A. \emph{Nanoscale} \textbf{2017}, \emph{9},
  16437--16445\relax
\mciteBstWouldAddEndPuncttrue
\mciteSetBstMidEndSepPunct{\mcitedefaultmidpunct}
{\mcitedefaultendpunct}{\mcitedefaultseppunct}\relax
\EndOfBibitem
\bibitem[Wang \latin{et~al.}(2018)Wang, Ying, Zhou, de~Vreede, Liu, and
  Tian]{wang2018fabrication}
Wang,~Y.; Ying,~C.; Zhou,~W.; de~Vreede,~L.; Liu,~Z.; Tian,~J. \emph{Scientific
  reports} \textbf{2018}, \emph{8}, 1234\relax
\mciteBstWouldAddEndPuncttrue
\mciteSetBstMidEndSepPunct{\mcitedefaultmidpunct}
{\mcitedefaultendpunct}{\mcitedefaultseppunct}\relax
\EndOfBibitem
\bibitem[Ying \latin{et~al.}(2018)Ying, Houghtaling, Eggenberger, Guha,
  Nirmalraj, Awasthi, Tian, and Mayer]{ying2018formation}
Ying,~C.; Houghtaling,~J.; Eggenberger,~O.~M.; Guha,~A.; Nirmalraj,~P.;
  Awasthi,~S.; Tian,~J.; Mayer,~M. \emph{ACS nano} \textbf{2018}, \emph{12},
  11458--11470\relax
\mciteBstWouldAddEndPuncttrue
\mciteSetBstMidEndSepPunct{\mcitedefaultmidpunct}
{\mcitedefaultendpunct}{\mcitedefaultseppunct}\relax
\EndOfBibitem
\bibitem[Arcadia \latin{et~al.}(2017)Arcadia, Reyes, and
  Rosenstein]{arcadia2017situ}
Arcadia,~C.~E.; Reyes,~C.~C.; Rosenstein,~J.~K. \emph{ACS nano} \textbf{2017},
  \emph{11}, 4907--4915\relax
\mciteBstWouldAddEndPuncttrue
\mciteSetBstMidEndSepPunct{\mcitedefaultmidpunct}
{\mcitedefaultendpunct}{\mcitedefaultseppunct}\relax
\EndOfBibitem
\bibitem[Briggs \latin{et~al.}(2015)Briggs, Charron, Kwok, Le, Chahal,
  Bustamante, Waugh, and Tabard-Cossa]{briggs2015kinetics}
Briggs,~K.; Charron,~M.; Kwok,~H.; Le,~T.; Chahal,~S.; Bustamante,~J.;
  Waugh,~M.; Tabard-Cossa,~V. \emph{Nanotechnology} \textbf{2015}, \emph{26},
  084004\relax
\mciteBstWouldAddEndPuncttrue
\mciteSetBstMidEndSepPunct{\mcitedefaultmidpunct}
{\mcitedefaultendpunct}{\mcitedefaultseppunct}\relax
\EndOfBibitem
\bibitem[Yanagi \latin{et~al.}(2018)Yanagi, Hamamura, Akahori, and
  Takeda]{yanagi2018two}
Yanagi,~I.; Hamamura,~H.; Akahori,~R.; Takeda,~K.-i. \emph{Scientific reports}
  \textbf{2018}, \emph{8}\relax
\mciteBstWouldAddEndPuncttrue
\mciteSetBstMidEndSepPunct{\mcitedefaultmidpunct}
{\mcitedefaultendpunct}{\mcitedefaultseppunct}\relax
\EndOfBibitem
\bibitem[Dissado \latin{et~al.}(1984)Dissado, Fothergill, Wolfe, and
  Hill]{dissado1984weibull}
Dissado,~L.; Fothergill,~J.; Wolfe,~S.; Hill,~R. \emph{IEEE Transactions on
  electrical insulation} \textbf{1984}, 227--233\relax
\mciteBstWouldAddEndPuncttrue
\mciteSetBstMidEndSepPunct{\mcitedefaultmidpunct}
{\mcitedefaultendpunct}{\mcitedefaultseppunct}\relax
\EndOfBibitem
\bibitem[Degraeve \latin{et~al.}(1995)Degraeve, Groeseneken, Bellens, Depas,
  and Maes]{degraeve1995consistent}
Degraeve,~R.; Groeseneken,~G.; Bellens,~R.; Depas,~M.; Maes,~H.~E. A consistent
  model for the thickness dependence of intrinsic breakdown in ultra-thin
  oxides. Electron Devices Meeting, 1995. IEDM'95., International. 1995; pp
  863--866\relax
\mciteBstWouldAddEndPuncttrue
\mciteSetBstMidEndSepPunct{\mcitedefaultmidpunct}
{\mcitedefaultendpunct}{\mcitedefaultseppunct}\relax
\EndOfBibitem
\bibitem[Peck and Zierdt(1974)Peck, and Zierdt]{peck1974reliability}
Peck,~D.~S.; Zierdt,~C. \emph{Proceedings of the IEEE} \textbf{1974},
  \emph{62}, 185--211\relax
\mciteBstWouldAddEndPuncttrue
\mciteSetBstMidEndSepPunct{\mcitedefaultmidpunct}
{\mcitedefaultendpunct}{\mcitedefaultseppunct}\relax
\EndOfBibitem
\bibitem[Berman(1981)]{berman1981time}
Berman,~A. Time-zero dielectric reliability test by a ramp method. Reliability
  Physics Symposium, 1981. 19th Annual. 1981; pp 204--209\relax
\mciteBstWouldAddEndPuncttrue
\mciteSetBstMidEndSepPunct{\mcitedefaultmidpunct}
{\mcitedefaultendpunct}{\mcitedefaultseppunct}\relax
\EndOfBibitem
\bibitem[Lloyd \latin{et~al.}(2005)Lloyd, Liniger, and Shaw]{lloyd2005simple}
Lloyd,~J.; Liniger,~E.; Shaw,~T. \emph{Journal of Applied Physics}
  \textbf{2005}, \emph{98}, 084109\relax
\mciteBstWouldAddEndPuncttrue
\mciteSetBstMidEndSepPunct{\mcitedefaultmidpunct}
{\mcitedefaultendpunct}{\mcitedefaultseppunct}\relax
\EndOfBibitem
\bibitem[McPherson(2010)]{mcpherson2010reliability}
McPherson,~J.~W. \emph{Reliability physics and engineering}; Springer,
  2010\relax
\mciteBstWouldAddEndPuncttrue
\mciteSetBstMidEndSepPunct{\mcitedefaultmidpunct}
{\mcitedefaultendpunct}{\mcitedefaultseppunct}\relax
\EndOfBibitem
\bibitem[Strong \latin{et~al.}(2009)Strong, Wu, Vollertsen, Sune, La~Rosa,
  Sullivan, and Rauch~III]{strong2009reliability}
Strong,~A.~W.; Wu,~E.~Y.; Vollertsen,~R.-P.; Sune,~J.; La~Rosa,~G.;
  Sullivan,~T.~D.; Rauch~III,~S.~E. \emph{Reliability wearout mechanisms in
  advanced CMOS technologies}; John Wiley \& Sons, 2009; Vol.~12\relax
\mciteBstWouldAddEndPuncttrue
\mciteSetBstMidEndSepPunct{\mcitedefaultmidpunct}
{\mcitedefaultendpunct}{\mcitedefaultseppunct}\relax
\EndOfBibitem
\bibitem[McPherson and Mogul(1998)McPherson, and
  Mogul]{mcpherson1998underlying}
McPherson,~J.; Mogul,~H. \emph{Journal of Applied Physics} \textbf{1998},
  \emph{84}, 1513--1523\relax
\mciteBstWouldAddEndPuncttrue
\mciteSetBstMidEndSepPunct{\mcitedefaultmidpunct}
{\mcitedefaultendpunct}{\mcitedefaultseppunct}\relax
\EndOfBibitem
\bibitem[McPherson \latin{et~al.}(2003)McPherson, Kim, Shanware, and
  Mogul]{mcpherson2003thermochemical}
McPherson,~J.; Kim,~J.; Shanware,~A.; Mogul,~H. \emph{Applied Physics Letters}
  \textbf{2003}, \emph{82}, 2121--2123\relax
\mciteBstWouldAddEndPuncttrue
\mciteSetBstMidEndSepPunct{\mcitedefaultmidpunct}
{\mcitedefaultendpunct}{\mcitedefaultseppunct}\relax
\EndOfBibitem
\bibitem[McPherson \latin{et~al.}(1998)McPherson, Reddy, Banerjee, and
  Le]{mcpherson1998comparison}
McPherson,~J.; Reddy,~V.; Banerjee,~K.; Le,~H. Comparison of E and 1/E TDDB
  models for SiO/sub 2/under long-term/low-field test conditions. Electron
  Devices Meeting, 1998. IEDM'98. Technical Digest., International. 1998; pp
  171--174\relax
\mciteBstWouldAddEndPuncttrue
\mciteSetBstMidEndSepPunct{\mcitedefaultmidpunct}
{\mcitedefaultendpunct}{\mcitedefaultseppunct}\relax
\EndOfBibitem
\bibitem[Yanagi \latin{et~al.}(2014)Yanagi, Akahori, Hatano, and
  Takeda]{yanagi2014fabricating}
Yanagi,~I.; Akahori,~R.; Hatano,~T.; Takeda,~K.-i. \emph{Scientific reports}
  \textbf{2014}, \emph{4}, 5000\relax
\mciteBstWouldAddEndPuncttrue
\mciteSetBstMidEndSepPunct{\mcitedefaultmidpunct}
{\mcitedefaultendpunct}{\mcitedefaultseppunct}\relax
\EndOfBibitem
\bibitem[Yamazaki \latin{et~al.}(2018)Yamazaki, Hu, Zhao, and
  Wanunu]{yamazaki2018photothermally}
Yamazaki,~H.; Hu,~R.; Zhao,~Q.; Wanunu,~M. \emph{ACS nano} \textbf{2018},
  \emph{12}, 12472--12481\relax
\mciteBstWouldAddEndPuncttrue
\mciteSetBstMidEndSepPunct{\mcitedefaultmidpunct}
{\mcitedefaultendpunct}{\mcitedefaultseppunct}\relax
\EndOfBibitem
\bibitem[Bandara \latin{et~al.}(2019)Bandara, Karawdeniya, and
  Dwyer]{bandara2019push}
Bandara,~Y. N.~D.; Karawdeniya,~B.~I.; Dwyer,~J.~R. \emph{ACS Omega}
  \textbf{2019}, \emph{4}, 226--230\relax
\mciteBstWouldAddEndPuncttrue
\mciteSetBstMidEndSepPunct{\mcitedefaultmidpunct}
{\mcitedefaultendpunct}{\mcitedefaultseppunct}\relax
\EndOfBibitem
\bibitem[Kowalczyk \latin{et~al.}(2011)Kowalczyk, Grosberg, Rabin, and
  Dekker]{kowalczyk2011modeling}
Kowalczyk,~S.~W.; Grosberg,~A.~Y.; Rabin,~Y.; Dekker,~C. \emph{Nanotechnology}
  \textbf{2011}, \emph{22}, 315101\relax
\mciteBstWouldAddEndPuncttrue
\mciteSetBstMidEndSepPunct{\mcitedefaultmidpunct}
{\mcitedefaultendpunct}{\mcitedefaultseppunct}\relax
\EndOfBibitem
\bibitem[Larkin \latin{et~al.}(2017)Larkin, Henley, Jadhav, Korlach, and
  Wanunu]{larkin2017length}
Larkin,~J.; Henley,~R.~Y.; Jadhav,~V.; Korlach,~J.; Wanunu,~M. \emph{Nature
  nanotechnology} \textbf{2017}, \emph{12}, 1169\relax
\mciteBstWouldAddEndPuncttrue
\mciteSetBstMidEndSepPunct{\mcitedefaultmidpunct}
{\mcitedefaultendpunct}{\mcitedefaultseppunct}\relax
\EndOfBibitem
\bibitem[Gilboa \latin{et~al.}(2018)Gilboa, Zrehen, Girsault, and
  Meller]{gilboa2018optically}
Gilboa,~T.; Zrehen,~A.; Girsault,~A.; Meller,~A. \emph{Scientific reports}
  \textbf{2018}, \emph{8}, 9765\relax
\mciteBstWouldAddEndPuncttrue
\mciteSetBstMidEndSepPunct{\mcitedefaultmidpunct}
{\mcitedefaultendpunct}{\mcitedefaultseppunct}\relax
\EndOfBibitem
\bibitem[Okada \latin{et~al.}(2007)Okada, Ota, Nabatame, and
  Toriumi]{okada2007dielectric}
Okada,~K.; Ota,~H.; Nabatame,~T.; Toriumi,~A. Dielectric breakdown in high-K
  gate dielectrics-Mechanism and lifetime assessment. Reliability physics
  symposium, 2007. proceedings. 45th annual. ieee international. 2007; pp
  36--43\relax
\mciteBstWouldAddEndPuncttrue
\mciteSetBstMidEndSepPunct{\mcitedefaultmidpunct}
{\mcitedefaultendpunct}{\mcitedefaultseppunct}\relax
\EndOfBibitem
\bibitem[D’Aprano \latin{et~al.}(1996)D’Aprano, Salomon, and
  Iammarino]{daprano1996conductance}
D’Aprano,~A.; Salomon,~M.; Iammarino,~M. \emph{Journal of electroanalytical
  chemistry} \textbf{1996}, \emph{403}, 245--249\relax
\mciteBstWouldAddEndPuncttrue
\mciteSetBstMidEndSepPunct{\mcitedefaultmidpunct}
{\mcitedefaultendpunct}{\mcitedefaultseppunct}\relax
\EndOfBibitem
\bibitem[Plesa and Dekker(2015)Plesa, and Dekker]{plesa2015data}
Plesa,~C.; Dekker,~C. \emph{Nanotechnology} \textbf{2015}, \emph{26},
  084003\relax
\mciteBstWouldAddEndPuncttrue
\mciteSetBstMidEndSepPunct{\mcitedefaultmidpunct}
{\mcitedefaultendpunct}{\mcitedefaultseppunct}\relax
\EndOfBibitem
\end{mcitethebibliography}
\end{document}